\documentclass[aps,prl,preprint,groupedaddress,showpacs]{revtex4}
\usepackage{graphicx}
\usepackage{graphics}
\usepackage{epsfig}	
\usepackage{bm}

\begin{document}

\title{Mass functions in coupled Dark Energy models.}
\vglue .4truecm

\author{ Roberto Mainini, Silvio Bonometto}
\vglue .1truecm

\affiliation{Department of Physics G.~Occhialini -- Milano--Bicocca
 University, Piazza della Scienza 3, 20126 Milano, Italy }
 \affiliation{ I.N.F.N., Sezione di Milano}
\date{\today}

\begin{abstract}
We evaluate the mass function of virialized halos, by using Press \&
Schechter (PS) and/or Steth \& Tormen (ST) expressions, for
cosmologies where Dark Energy (DE) is due to a scalar
self--interacting field, coupled with Dark Matter (DM). We keep to
coupled DE (cDE) models known to fit linear observables. To implement
the PS--ST approach, we start from reviewing and extending the results
of a previous work on the growth of a spherical top--hat fluctuation
in cDE models, confirming their most intriguing astrophysical feature,
{\it i.e.} a significant baryon--DM segregation, occurring well before
the onset of any hydrodynamical effect. Accordingly, the predicted
mass function depends on how halo masses are measured. For any option,
however, the coupling causes a distortion of the mass function, still
at $z=0$. Furthermore, the $z$--dependence of cDE mass functions is
mostly displaced, in respect to $\Lambda$CDM, in the opposite way of
uncoupled dynamical DE. This is an aspect of the basic underlying
result, that even a little DM--DE coupling induces relevant
modifications in the non--linear evolution. Therefore, without causing
great shifts in linear astrophysical observables, the DM--baryon
segregation induced by the coupling can have an impact on a number of
cosmological problems, {\it e.g.}, galaxy satellite abundance, spiral
disk formation, apparent baryon shortage, entropy input in clusters,
etc..

\end{abstract}

\maketitle
\section{Introduction}
\label{sec:intro}
A first evidence of Dark Energy (DE) came from the Hubble diagram of
SNIa, showing an accelerated cosmic expansion, but a {\it flat}
cosmology with $\Omega_{m} \simeq 0.25$, $h \simeq 0.73$ and
$\Omega_{b} \simeq 0.042$ is now required by CMB and LSS observations
and this implies that the gap between $\Omega_m$ and unity is to be
filled by a smooth non--particle component, whose nature is one of the
main puzzles of cosmology. ($\Omega _{m,b}$: matter, baryon density
parameters; $h$: Hubble parameter in units of 100 km/s/Mpc; CMB:
cosmic microwave background; LSS: large scale structure.)

If DE is a false vacuum, its pressure/density ratio $w =
p_{DE}/\rho_{DE} $ is strictly -1. This option, however, implies a
severe fine tuning at the end of the electroweak transition. Otherwise
DE could be a scalar field $\phi$, self--interacting through a
potential $V(\phi)$ \cite{sugra}, \cite{RP}, so that
\begin{equation}
\rho_{DE}=\rho_{k,DE}+\rho_{p,DE} \equiv {{\dot{\phi }}^{2}/2a^{2}}+V(\phi ),~
~~~~
p_{DE}=\rho_{k,DE}-\rho_{p,DE}~,
\end{equation}
provided that dynamical equations yield $\rho _{k,DE}/V\ll 1/2$, so
that $-1/3 \gg w>-1$. Here
\begin{equation}
ds^{2} = a^{2}(\tau) (-d\tau^{2}+ dx_i dx^i ) ~,~~~~~ (i=1,..,3)
\end{equation}
is the background metric and dots indicate differentiation with
respect to $\tau $ (conformal time). This kind of DE is dubbed {\it
dynamical} DE (dDE) or {\it quintessence}; the $w$ ratio then exhibits
a time dependence set by the shape of $V(\phi)$. Much work has been
done on dDE (see, e.g., \cite{PR} and references therein), also aiming
at restricting the range of acceptable $w(\tau)$'s, so gaining an
observational insight onto the physics responsible for the potential
$V(\phi)$.

As a matter of fact, the dark cosmic components are one of the most
compelling evidences of physics beyond the standard model of
elementary interactions and, while lab experiments safely exclude
non--gravitational baryon--DE interactions, DM--DE interactions are
constrained just by cosmological observations. In turn, DM--DE
interactions could ease the cosmic coincidence problem
\cite{Amendola1999}, i.e. that DM and DE densities, differing by
orders of magnitude since ever, approach one another at today's
eve. 

In a number of papers, constraints on coupling, coming from CMB and
LSS observations, were discussed \cite{Amendola1999,lucaclaudia,
maccio}. This note aims at formulating predictions on the mass
function of bound systems in coupled DE (cDE) cosmologies, by using
Sheth \& Tormen \cite{ST} expressions, known to improve the original
Press \& Schechter \cite{PS} approach. Our final scope amounts to
finding stronger constraints on DM--DE coupling, arising from a
comparison of observational data with our predictions, opening a basic
window on the nature and origin of these very components. Our analysis
will be restricted to SUGRA \cite{brax} and RP (Ratra--Peebles) \cite{RP} potentials
$$
SUGRA ~~~~~~~~~~~~~~~~~~~~~~~~~~~~~~
V(\phi) = (\Lambda^{\alpha+4}/\phi^\alpha) \exp(4 \pi \phi^2/m_p^2)
$$
\begin{equation}
\label{ptntl}
RP ~~~~~~~~~~~~~~~~~~~~~~~~~~~~~~~~~~~~~~~~~~~~~~~~~~
V(\phi) = (\Lambda^{\alpha+4}/\phi^\alpha)
\end{equation}
($m_p=G^{-1/2}$: Planck mass), admitting tracker solutions. This will
however enable us to focus on precise peculiarities, not caused by the
shape of $V(\phi)$ but by the coupling itself. Let us also remind
that, once the DE density parameter $\Omega_{DE}$ is assigned, either
$\alpha$ or the energy scale $\Lambda$, in the potentials
(\ref{ptntl}), can still be freely chosen. In this paper we show
results for $\Lambda = 10^2\, $GeV; minor quantitative shifts occur
when varying $\log_{10}(\Lambda/{\rm GeV})$ in the 1--4 range. The RP
potential will be mostly considered to test the effects of varying DE
nature.

The effects of coupling can be seen in the background equations for DE
and DM, reading
\begin{equation}
\ddot{\phi} + 2(\dot a/a) \dot{\phi} + a^2 V_{,\phi} = 
\sqrt{16\pi G/3} \beta a^{2} \rho _{c}~, 
~~~~ \dot{\rho _{c}} + 3 (\dot a/a)\rho_{c}  =  
-\sqrt{16\pi G/3} \beta \rho _{c} \dot \phi~;
\label{backg}
\end{equation}
here $\beta$ sets the coupling strength and, all through this paper,
we take it constant (and, in particular, independent of $\phi$; a
different case, which can be physically significant and may deserve a
separate treatment \cite{axion}) with values $\beta=0.05$ or 0.20$\,
$. In previous work this was considered a {\it small} strength. CMB
data set a limit $\beta <\sim 0.6$; a stronger constraint found by
\cite{maccio}, by studying cluster profiles in n--body simulations,
applies to RP potentials only.

On the contrary, even for the smaller coupling we considered, we find
significant deviations from uncoupled models in non--linear
observables. Such deviations exhibit intriguing features, suggesting
possible ways out from a number of astrophysical problems, while the
small coupling strength scarcely affects linear observables.

As far as the other model parameters are concerned, we take density
parameters $\Omega_m=0.25$, $\Omega_b=0.042$, $\Omega_c=0.208$ and
$h=0.73$. This choice approaches the best fit of a $\Lambda$CDM model
to available data. A best fit to cDE models will certainly yield
slightly different parameters, but this implies just minor shifts in
our quantitative findings.

\section{Baryon and Dark Matter dynamics}
The essential novel feature induced by DM--DE coupling in non--linear
structures is baryon--DM segregation. This was shown in \cite{mainini}
(paper I, hereafter). The reason why segregation occurs can be easily
illustrated by considering, aside of the eq.~(\ref{backg}), ruling
background dynamics, the couple of equations telling us how the baryon
(DM) density fluctuations $\delta_b$ ($\delta_c$) and velocity fields
\begin{equation}
\theta_{c,b}=i{\frac{\bf k \cdot {\bf v_{c,b}}}{{\cal H}}}.
\label{theta}
\end{equation}
depend on $\tau$ (${\cal H} = \dot a/a$; {\bf k}: wavenumber of the
fluctuation considered). They read
\begin{eqnarray}
{\delta _{c}}\, '' & = & 
-{\delta _{c}}\, '\big (1+\frac{{\mathcal{H}}'}
{\mathcal{H}}-2\beta X\big )+\frac{3}{2}
(1+\frac{4}{3}\beta ^{2})\Omega _{c}\delta _{c}+
\frac{3}{2}\Omega _{b}\delta _{b},\nonumber\\
{\delta _{b}}\, '' & = & 
-{\delta _{b}}\, '\big (1+\frac{{\mathcal{H}}'}
{\mathcal{H}}\big )+\frac{3}{2}(\Omega _{c}\delta_{c}
+\Omega _{b}\delta _{b}),\label{delta2}
\end{eqnarray}
\begin{eqnarray}
{\theta _{c}}\, ' & = & 
-\theta _{c}\big (1+\frac{{\mathcal{H}}'}{\mathcal{H}}
-2\beta X\big )-\frac{3}{2}\big (1+\frac{4}{3}\beta^{2}\big)
\Omega _{c}\delta _{c}-
\frac{3}{2}\Omega _{b}\delta _{b},
\nonumber,
\\
{\theta _{b}}\, ' & = & 
-\theta _{b}\big( 1+\frac{{\mathcal{H}}'}{\mathcal{H}} \big)
-\frac{3}{2}(\Omega _{c}\delta _{c}+\Omega _{b}\delta _{b})~.
\label{ar2}
\end{eqnarray}
Here $'$ yields differentiation with respect to $\alpha = \ln a$ and
\begin{equation}
 X = \sqrt{4 \pi/3}\, \dot \phi/(m_p {\cal H})~.
\label{ics}
\end{equation}
Both eqs.~(\ref{backg}) and
eqs.~(\ref{delta2})--(\ref{ar2}) show that,
as soon as $\beta \neq 0$, baryons and DM have different
dynamics. Neglecting radiation and any hydrodynamical effects, baryons
move along geodesics. On the contrary, DM particles feel also
non--gravitational forces and, as it happens in the presence of any
non--gravitational interaction, do not follow geodesics. This is why
baryon--DM segregation occurs.

Starting from eqs.~(\ref{delta2})--(\ref{ar2}), the evolution of
linear fluctuations in cDE cosmologies was studied by
\cite{Amendola1999}. Here we shall put ourselves into a different
physical context, by lifting the restriction $\delta_{c,b} \ll 1$, but
considering just cases when a full relativistic treatment is
unessential. Then, dealing with scales well below horizon and with
non--relativistic particles, the long--range force carried by the DE
field $\phi$ can be described through corrections to {\it newtonian}
gravity. They amount to assuming:

\noindent
(i) DM particle masses to vary, so that
\begin{equation}
M_{c} (\tau) = M_{c} (\tau_i) \exp[-C (\phi - \phi_i)].
\label{masva}
\end{equation}

\noindent
(ii) The gravitational constant between DM particles to become $G^* =
\gamma G$.

\noindent
Here
\begin{equation}
C = \sqrt{16\pi G/3} ~\beta ~~,~~~~
\gamma = 1+4\beta^2/3
\label{Cgamma}
\end{equation}
A proof is given in Appendix A, following \cite{Amendola03} and
\cite{maccio}.

A newtonian treatment is suitable to study the growth of spherical
{\it top--hat} fluctuations in the DM and baryon components and to
show how their segregation occurs in the relevant physical cases. In
fact, non--linearity starts well after matter--radiation decoupling
and the top scales to be considered, galaxy cluster scales, lay well
below the horizon. 

The evolution of fluctuations in this regime was debated in paper
I. Here we shall use its results to understand the shaping of mass
functions. We therefore start from reviewing them and complementing
them with new quantitative outputs.

\section{Spherical top--hat evolution }
Although in all cosmologies, apart sCDM, the system of equations
ruling the spherical growth requires a numerical solution, in cDE
models the numerical approach is far more essential. In fact, in all
cases apart cDE, the only variable describing a {\it top--hat}
fluctuation is its radius $R$. It is assumed to expand, initially, at
the same rate of the scale factor $a$, obeying an equation similar to
the scale factor in a closed model. The greater density inside the top
hat slows then down the increase rates of $R$ in respect to $a$, so
that the inner density $\rho(<R)$, although decreasing, becomes
greater and greater than the average density $\rho$. Eventually, at a
time $t_{ta}$, when the density contrast $\Delta = \rho(<R)/\rho$
attains a suitable value $\chi$, $R$ stops and starts decreasing. The
equation it obeys would then cause $R$ to vanish within a finite time
$t_c$. In sCDM cosmologies, $\chi = (3\pi/4)^2$ and $t_c/t_{ta} =
2$. In other non--cDE cosmologies, $\chi$ and the ratio $t_c/t_{ta} $
take just slightly different values.

After $t_{ta}$, however, the very density inside $R$ is increasing, so
that, in any realistic fluctuation, unless radiation disposes the heat
produced by the $p\, dV$ work, virial equilibrium is attained and the
collapse essentially stops when the fluctuation radius is $R_v$. In a
sCDM cosmology this occurs for $R_v/R_{ta} = 1/2$. Taking onto account
the symultaneous growth of $a$, the virial density contrast is then
$\Delta_v = 32\chi \simeq 180$. In other cosmologies the $R_v/R_{ta}$
ratio is slightly different from 1/2 and $\Delta_v$ is mostly above
100, but, in most cases, significantly below 180. A usual assumption
is that the time when the system is stabilized into a virialized
configuration coincides with $t_c$. Let us also remind that while, in
sCDM, $\Delta_v$ is independent from $t_c$, this is no longer true in
all models where the density ratio between DE and material components
depends on time.

Quantitative results for the spherically symmetric growth in
$\Lambda$CDM models were given by \cite{Lahav_brian}; their extension
to models with uncoupled dDE is due to \cite{io,collapse1}. In these
cosmologies, the simple system of equations yielding spherical
evolution requires a numerical solution.

Let us now consider the same problem in the cDE case. This problem had
been considered in \cite{collapse2}, but its approximated treatment
did not allow the authors to focus the main physical effects.
Following \cite{mainini}, let us outline, first of all, that linear
fluctuations in baryons and DM are already different \cite{Amendola03}
and have a ratio
\begin{equation}
{\delta_{b} \over \delta_c }
\simeq
{ 3 \Omega_c \over
3 \gamma\, \Omega_c + 4\beta\, X \mu}~.
\label{biasl}
\end{equation}
Here $X$ is given in eq.~(\ref{ics}), $\gamma$ is given in
eq.~(\ref{Cgamma}); $\mu = (\dot \delta_{c,b}/\delta_{c,b})/{\cal
H}$. When considering a spherical top--hat fluctuation, whose initial
radius $R_{TH,i}=R_b=R_c$, eq.~(\ref{biasl}) sets the initial ratio
between DM and baryon fluctuation amplitudes.

Initial conditions are set so that both $R_b$ and $R_c$ initially grow
as the scale factor. However, because of the different interaction
strength, as soon as non--linear effects become significant, $R_{b}$
starts exceeding $R_{c}$. Hence, a part of the baryons initially
within $R_{TH,i}$ leak out from $R_c$, so that DM does not feel the
gravity of all baryons, while baryons above $R_c$ feel the gravity
also of initially unperturbed DM layers. As a consequence, above
$R_c$, the baryon component deviates from a {\it top--hat} geometry,
while a secondary perturbation in DM arises also above $R_c$ itself.

After reaching maximum expansion, contraction will also start at
different times for different components and layers. Then, inner
layers will approach virialization before outer layers, whose later
fall--out shall however perturb their virial equilibrium.  This
already outlines that the onset of virial equilibrium is a complex
process.

Furthermore, when the external baryon layers fall--out onto the
virialized core, richer of DM, they are accompanied by DM materials
originally outside the top--hat, perturbed by baryon over--expansion.

The time when the greatest amount of materials, originally belonging
to the fluctuation, are in virial equilibrium occurs when the DM
top--hat has virialized, together with the baryon fraction still below
$R_c$. A large deal of baryonic materials are then still falling
out. But, when they will accrete onto the DM--richer core, they will
not be alone, carrying with them originally alien materials.  There
will be no discontinuity in the fall--out process when all original
baryons are back. The infall of outer materials just attains then a
steady rate.


In order to follow the dynamical evolution of a systems where each
layer feels a different and substance--dependent force, a set of
concentric shells, granting a sufficient radial resolution, needs to
be considered. In the next section we shall provide the equations of
motion for Lagrangian shell radii and review the whole expansion and
recontraction dynamics.

\section{Time evolution of concentric shells}

\subsection{The spherical growth}
Dynamical equations can be written by using comoving radii $b_n =
R_b^n/a $ ($ c_n = R_c^n/a $) for the $n$--th baryon (DM) shell.  Let
$M^n_{b,c}$ be the masses of the $n$--th layer, $\bar M^{n}_{b,c}$ be
the masses in the same layer in the absence of perturbations, and
$\delta M^n_{b,c} = M^{n}_{b,c} - \bar M^{n}_{b,c}$ be the mass
excesses for the $n$--th layer. Baryon shells keep a constant mass in
the whole process, while
\begin{equation}
 M^n_c (\tau) = M^n_c (\tau_{in})
\exp\{-C[\phi(\tau)-\phi(\tau_{in})]\}~,
\label{mvar}
\end{equation}
according to eq.~(\ref{masva}). Similarly, let $\Delta M_{b,c}(<r)=
M_{b,c}(<r)- \bar M _{b.c}(<r)$ be the mass excess within the physical
radius $R=ar$ and let then be
\begin{equation}
\Delta M^n_b = \Delta M_{b}(<b_n) + \Delta M_{c}(<b_n)~,~~~
\Delta M^n_c = \Delta M_{b}(<c_n) + \gamma \Delta M_{c}(<c_n)~,
\label{dm}
\end{equation}
including a correction aimed to take into account, besides of the mass
variation (eq.~\ref{mvar}), also that $G^* \neq G$ in DM--DM
interactions.

A systematic discrepancy will arise between $b_n$ and $c_n$ and will
require taking into account only the fraction of the {\it last} shell
below the radius considered. In paper I it is shown that the shell
dynamics can then be described through the equations
\begin{equation}
\ddot c_n = - \left( {\dot a \over  a} - C\dot \phi \right) \dot c_n
- G {\Delta M^n_c \over a c_n^2}~,~~~
\ddot b_n = - {\dot a \over a} \dot b_n
- G {\Delta M^n_b \over a b_n^2}~,
\label{shells}
\end{equation}
to be integrated together with the first of eqs.~(\ref{backg}) and the
Friedman equation.

\begin{figure}
\begin{center}
\includegraphics*[width=12cm]{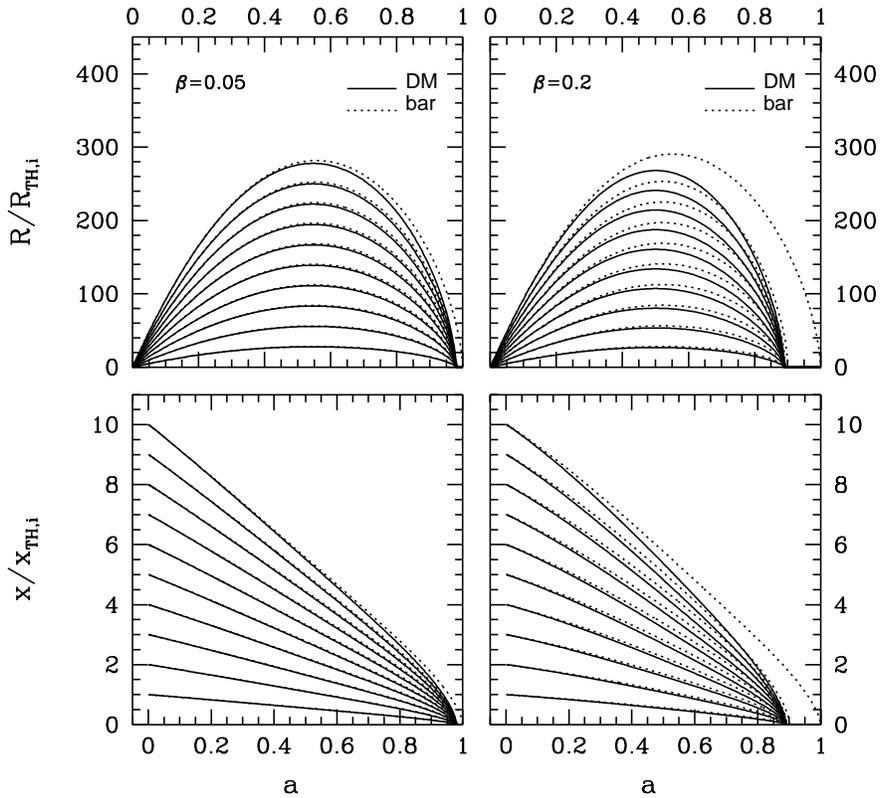}
\end{center}
\vskip -1.2truecm
\caption{Evolution of a sample of baryon and DM layer radii, extending
up to the top--hat radius; upper (lower) panels describe physical
(comoving) radii. In all plots the leaking out of the upper baryon
layer is clearly visible.}
\label{shell}
\end{figure}

Until $b_n \sim c_n$, it is $\Delta M^n_{b} < \Delta M^n_{c}$, for any
$\beta > 0$. Hence, the gravitational push felt by DM layers is
stronger. The extra term $C\dot \phi \dot c_n$ adds to this push. In
fact, the comoving variable $ c_n$ has however a negative derivative
(see the lower panels in Figure \ref{shell}). As a consequence, $c_n$
decreases more rapidly than $b_n$, so that the $n$--th baryon radius
may gradually exceed the $(n+1)$--th DM shell, etc.$\, $. As a
consequence, the sign of $\Delta M^n_{b} - \Delta M^n_{c}$ can even
invert and one can evaluate for which value of $\phi-\phi_i$ this
occurs. Once $\phi(\tau)$ is known, also the time when this occurs can
be found.  Until then, however, DM fluctuations expand more slowly and
mostly reach their turn--around point earlier, while baryons gradually
leak out from the fluctuation bulk.

\begin{figure}
\begin{center}
\includegraphics*[width=10cm]{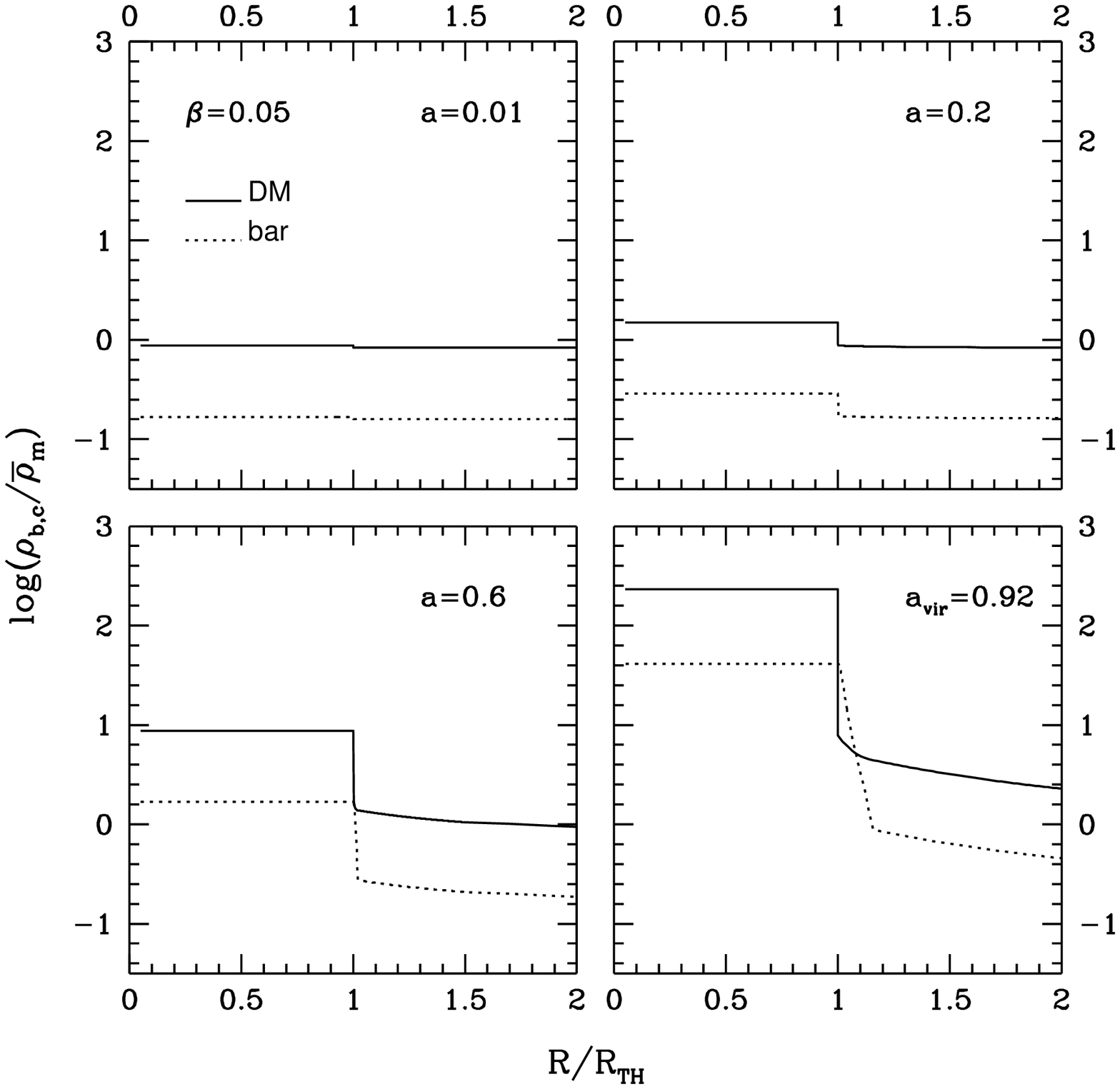}
\end{center}
\vskip -1.3truecm
\caption{Density profiles at different $a$ values for a cDE model with
$\beta = 0.05$. Solid (dotted) lines refer to DM (baryons).  Notice
the progressive deformation of the baryon profile (dotted lines) in
respect of the DM profile (solid line).}
\label{profili005}
\begin{center}
\includegraphics*[width=10cm]{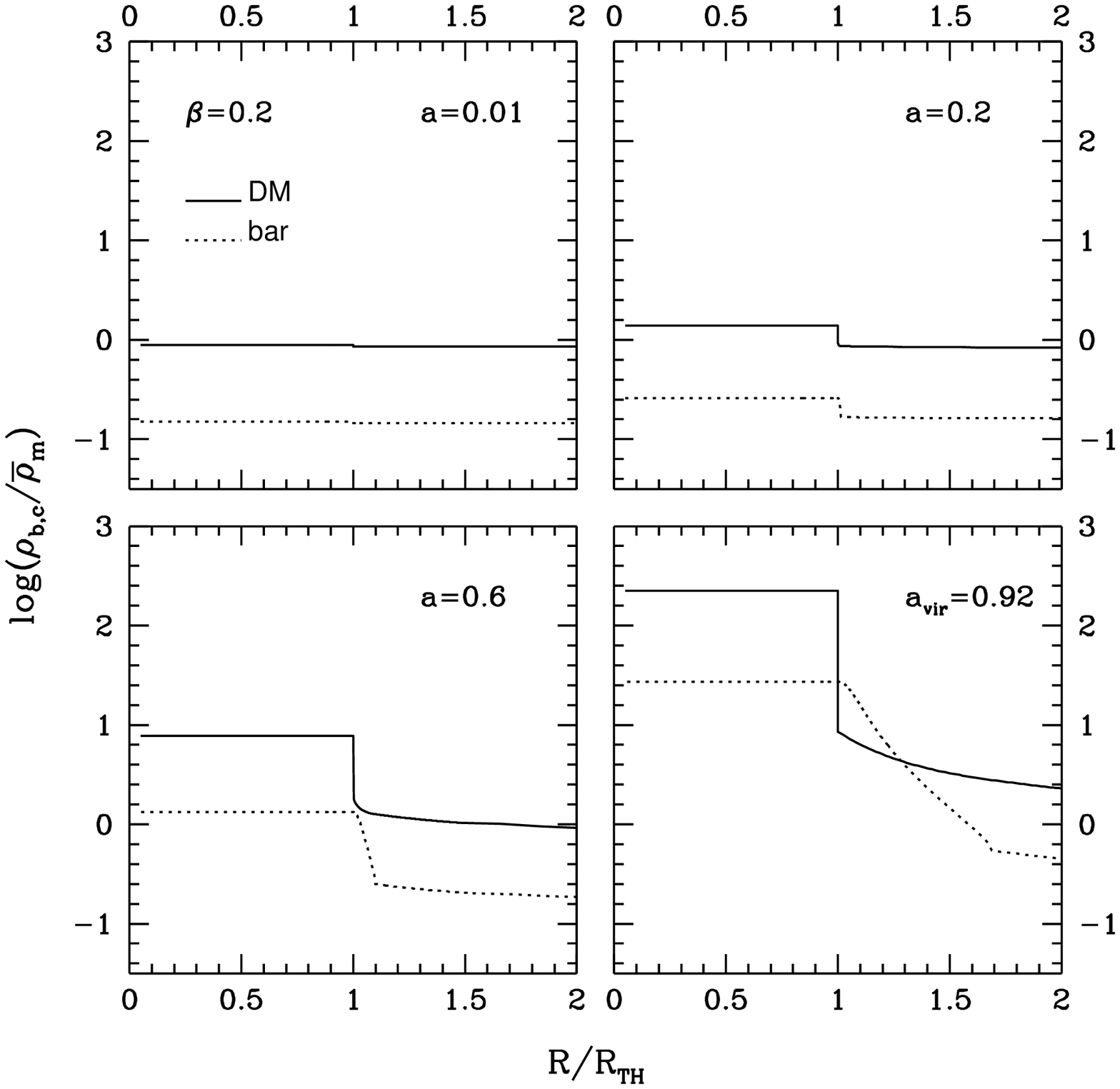}
\end{center}
\vskip -1.3truecm
\caption{Density profiles at different $a$ values for $\beta = 0.2$
model. The deformation of the baryon profile (dotted lines) in respect
of the DM profile (solid line) is stronger than in the previous
Figure.}
\label{profili02}
\end{figure}
The whole behavior is visible in the Figure \ref{shell}, for samples
of $R_{c,b}$ and $c$--$b$ (the ordinate label $\, $x$\, $ stands for
either $c$ or $b$). The greatest radii shown are the baryon top--hat
radii. Solid (dotted) lines yield the $a$ dependence for DM (baryon)
shell radii. The actual number of radii $R^n_{b,c}$ used in the
equations depend on the precision wanted; radii do not need to be
equi--spaced. To our aims (a $0.1\, \%$ precision), $\sim 1000$ radii
were sufficient. If more precision is needed, it can be easily
achieved at the expenses of using a greater computer time. With a
single processor 1.5 Gflops PC, our optimized program takes $\sim 60$
minutes to run a single case, with $\sim 1000$ radii.

The full collapse shown in Figure \ref{shell} is not expected in any
physical case, as the $R_n$ decrease stops when virialization is
attained. Figs.~\ref{profili005}, \ref{profili02} therefore assume
that the spherical growth stops when all DM originally in the
top--hat, and the baryons kept inside it, virialize.  This figure
describes the gradual deformation of the top--hat profile.  Already at
$a = 0.2$, well before the turn--around, the slope of the top--hat
boundary, for baryons, is no longer vertical.  At $a=0.6$
(approximately turn--around), not only the baryon boundary is bent,
but a similar effect is visible also for DM. The effect is even more
pronounced at $a_{vir } \simeq 0.92$. Figures similar to the upper
plots in Fig.~\ref{shell} and to Figs.~\ref{profili005},
\ref{profili02} were already shown in paper I, although for different
model parameters.

The increased density of DM shells outside the top--hat is relevant,
here, because it fastens the recollapse of outer baryons. In turn, the
enhanced baryon density, outside from DM top--hat, modifies the
dynamics of external DM layers, as well.

The most significant physical effect, however, is the outflow of
baryon layers from the DM top--hat. The outflown baryon fraction
increases with $a$. For $a \sim 0.92$, i.e. when DM and inner baryons
have attained their virialization radius, the fraction of baryons
which have leaked out from the fluctuation is so large as $\sim 10\,
\%$, even for $\beta = 0.05$; it reaches $\sim 40\, \%$ for $\beta =
0.2\, $. These values increase by an additional $ 10\, \%$ in the RP
case.

\subsection{Virialization}
In Figures  \ref{profili005} and \ref{profili0} we assumed the present time to coincide with
virialization for {\it all DM inside the top--hat fluctuation and all
baryons kept inside it}, excluding therefore a large deal of baryons,
either 10$\, $ or $40\, \%$, initially inside the top--hat, but
leaking out during the fluctuation growth.

It must also be outlined that much care ought to be taken to define
the virialization condition, for materials within any radius $R$,
reading
\begin{equation}
2\, T(R) = R\, dU(R)/dR~,
\label{viri}
\end{equation}
by summing up DM and baryon kinetic energies
\begin{equation}
T_c(R)= 2\pi \int_0^R dr\, r^2  \rho_c(r) ~ \dot r^2~,~~ 
T_b(R)= 2\pi \int_0^R dr\, r^2  \rho_b(r) ~ \dot r^2
\label{kinetic}
\end{equation}
and taking into account that, during the whole fluctuation evolution,
potential energies include three terms, due to self--interaction,
mutual interaction, interaction with the DE field. More in detail, for
DM and baryons, we have
\begin{equation}
U_c(R) = U_{cc}(R)+ U_{cb}(R)+ U_{c,DE}(R) = 4 \pi \int^R_0 dr ~r^2\,
\rho_c(r)~ [\bar \Psi_c(r) + \Psi_b(r) + \Psi_{DE}(r)]~,
\label{uc}
\end{equation}
\begin{equation}
U_b(R) = U_{bb}(R)+ U_{bc}(R)+ U_{b,DE}(R) =
4 \pi \int^R_0 dr ~r^2\, \rho_b(r)~ [\Psi_b(r) + \Psi_c(r) + \Psi_{DE}(r)]~,
\label{ub}
\end{equation}
respectively. While
\begin{equation}
\Psi_b (r) = -{4 \pi \over 3} G \rho_{b}(r) r^2~,~~
\Psi_{DE} (r) = -{4 \pi \over 3} G \rho_{DE}(r) r^2~,~~
\end{equation}
in both expressions, a subtle difference exists for $\Psi_c$.
In $U_b$ we have simply
\begin{equation}
\Psi_c (r) =  -{4 \pi \over 3} G \rho_{c}(r) r^2~,~~
\end{equation}
as for the other components, but this expression is different in
$U_c$, where it reads
\begin{equation}
\bar \Psi_c (r) =  -{4 \pi \over 3} \{ \gamma G [\rho_{c} - \bar \rho_c (r)] 
+ G \bar \rho_{c} \}r^2~,
\label{diversa}
\end{equation}
$\bar \rho_c$ being the background DM density. The different dynamical
effect of background and DM fluctuation arises from the different ways
how its interaction with DE is treated. Energy exchanges between DM
and DE, for the background, are accounted for by the r.h.s. terms in
eqs.~(\ref{backg}). In this case no newtonian approximation was
possible and was made. For DM fluctuation, instead, the effects of
DM--DE exchanges are described by a correction to the gravitational
constant $G$, becoming $G^* = \gamma G$, which adds to the dependence
of DM density on $\phi$. This ought to be taken into account in the
fluctuation evolution, as is done in eq.~(\ref{diversa}).

The above expressions hold for any $R$. However, if $R$ coincides with
DM top--hat (or is smaller than it), the densities $\rho_{b,c,DE}$ do
not depend on $r$. Densities depend on $r$ if one aims to take somehow
into account the shells where baryons, initially belonging to the
top--hat, have flown. In the former case, the integrals (\ref{uc}),
(\ref{ub}) and (\ref{kinetic}) can be easily performed and yield
\begin{equation}
U_c(R)= (3/5)G [M_c(M_b+\bar M_c) + \gamma M_c \Delta M_b]/R - (4 \pi/
5) M_c \rho_{DE} R_2
\end{equation}
\begin{equation}
U_b(R)= -(3 / 5)G[M_b (M_b +  M_c)]/ R
- (4 \pi /5) M_b \rho_{DE} R^2
\end{equation}
\begin{equation}
T_c(R)= (3/10)~M_c \dot R^2 
\end{equation}
where the relation $\dot r / r = \dot R / R$ is used to calculate
$T_c(R)$.  Note that this relation is not valid for $T_b(R)$ because
different baryon layers have different growth rates.  Kinetic energy
for baryons is then obtained by
\begin{equation}
T_b(R) = \sum_n T_b^n = \sum_n {1 \over 2} M_b^n (\dot R_b^n)^2 
\end{equation}
Here the sum is extended on all $R_b^n < R$.

If integrals are extended above the DM top--hat, recourse to numerical
computations is needed.

\section{Mass functions in ${\bf c}$DE theories}
The expected physical behavior of a top--hat fluctuation, including
non--linear features, can be compared with the evolution of that
fluctuation if we assume that linear equation however hold,
indipendently of the actual amplitude the fluctuation has reached.
Actual gravitation prescribes that fluctuations approaching unity
abandon the linear regime, slow down their expansion rate, reach
maximum expansion, turn--around and recontract, finally recollapsing
to nil. While this occur, we can formally assume that linear equation
still hold and seek the value $\delta_{rc}$ that {\it linear}
fluctuations would have at the time $\tau_{rc}$ when, according to
actual gravitation, they have recollapsed.

Let us also remind that the linear evolution does not affect
fluctuation amplitude distributions. Accordingly, if fluctuation
amplitudes are distributed in a Gaussian way, this does not change in
the linear regime. Therefore, at the time $\tau_{rc}$, we can
integrate on the distribution of fluctuation amplitudes for a given
scale, taking those $> \delta_{rc}$, so finding the probability that
an object has formed and virialized over such scale.

As already outlined in the previous section, full recollapse is not
expected to occur. The usual assumption is however that the time
running between the achievement of the virialization prescription and
the time formally required for full recollapse is taken by the
fluctuation to achieve a relaxed virial equilibrium configuration.
This is the basic pattern of the PS--ST approach, that we aim now to
apply to cDE cosmologies.
\begin{figure}
\begin{center}
\includegraphics*[width=11cm]{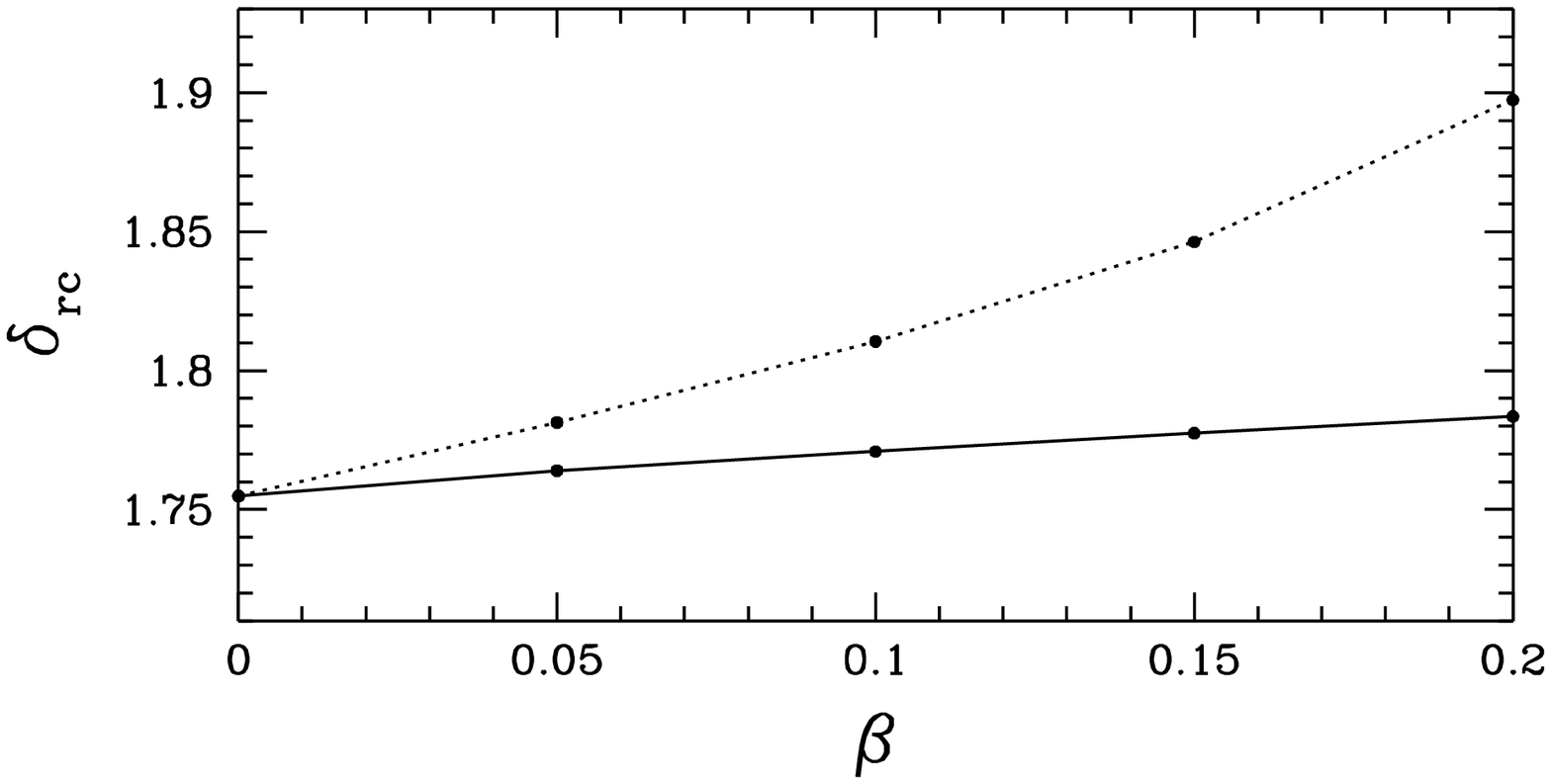}
\end{center}
\vskip -5.4truecm
\caption{$\beta$ dependence of $\delta_{c,rc}^{(c)}$ (dotted line) and
$\delta_{c,rc}^{(b)}\, $, at $z=0$ (solid line).  }
\label{deltac}
\begin{center}
\includegraphics*[width=11cm]{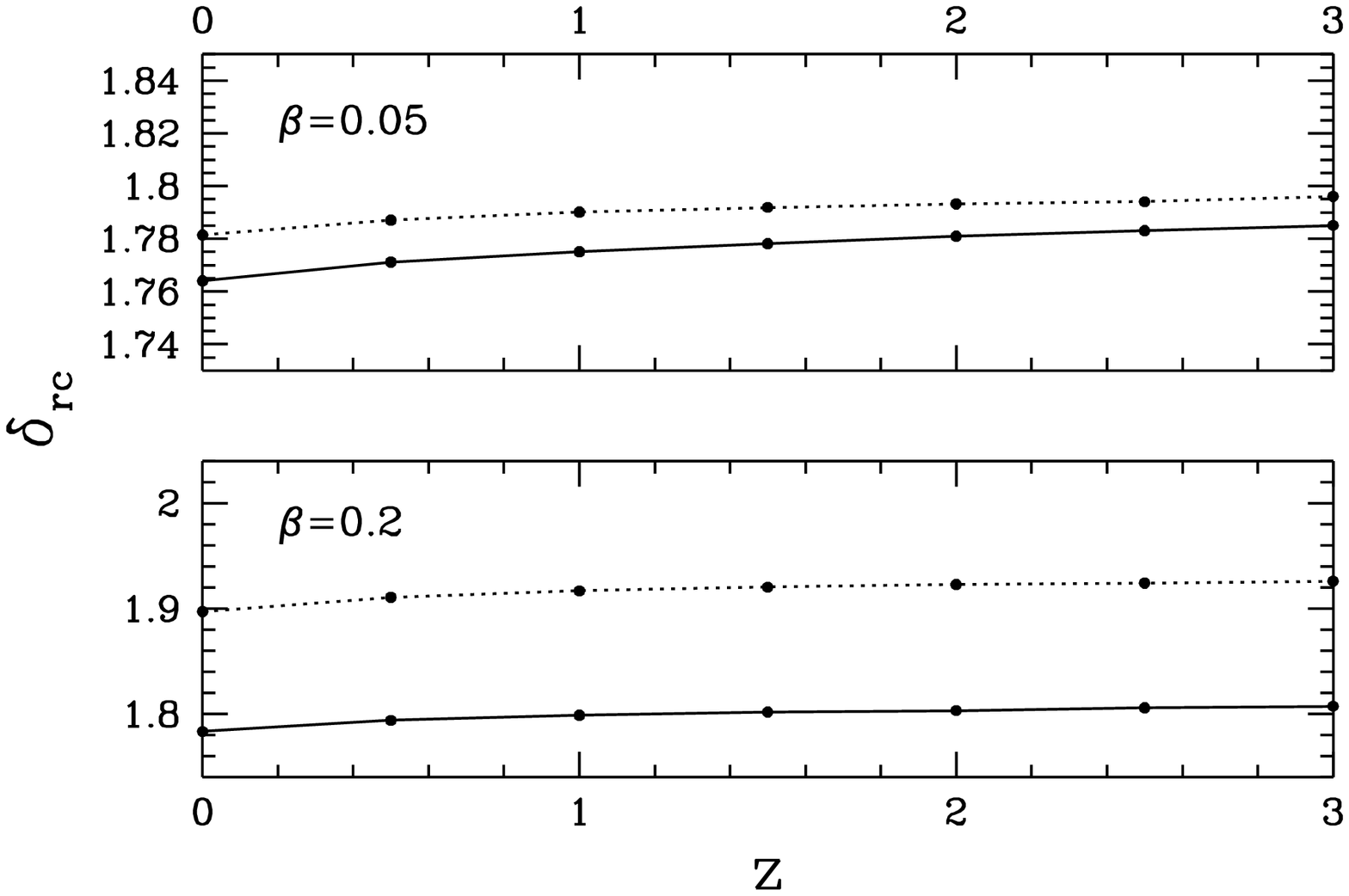}
\end{center}
\vskip -4.2truecm
\caption{Redshift dependence of $\delta_{c,rc}^{(c)}$ (dotted line)
and $\delta_{c,rc}^{(b)}\, $ (solid line).  
}
\label{dcz}
\end{figure}
\begin{figure}
\begin{center}
\includegraphics*[width=17cm,height=17cm]{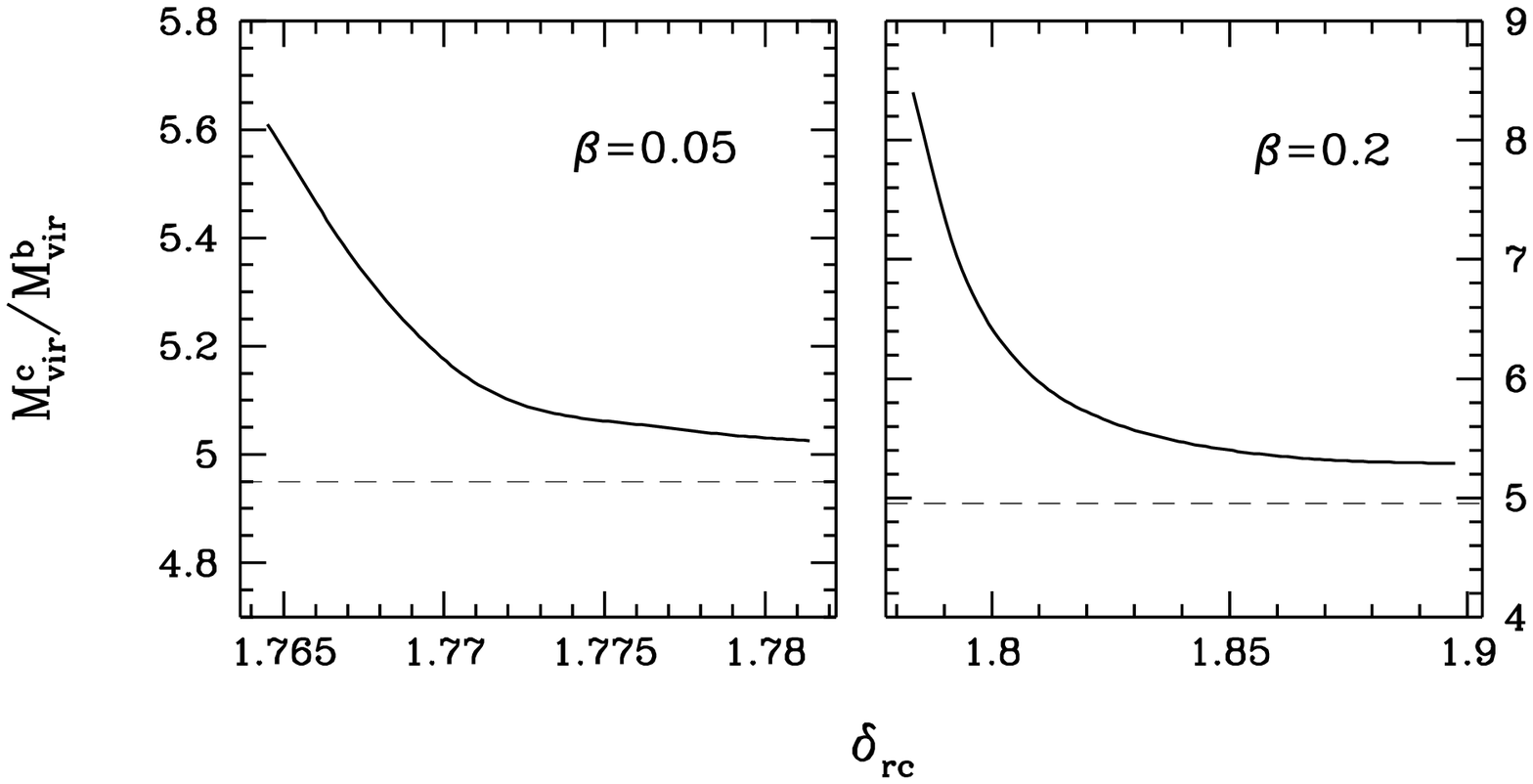}
\end{center}
\vskip -8.2truecm
\caption{Ratio between DM and baryon mass in virialized halos,
against the value used for $\delta_{c,rc}$ (DM masses are rescaled to the value 
they reach at $\tau_o$.  The dashed line
is the background $\Omega_c/\Omega_b$ ratio.
}
\label{mdelta}
\end{figure}
In the presence of DM--DE coupling, however, a novel feature must be
considered: starting from the initial amplitudes set by the linear
theory, DM and baryon fluctuations require different times to reach
full recollapse. Accordingly, if we require a baryon fluctuation to
reach full recollapse at the present time $\tau_o$ (or at any time
$\tau$), the {\it initial} linear $\delta_b$ and $\delta_c$ must have
been greater than those required to allow full recollapse at the
present time $\tau_o$ (or at any time $\tau$) for a DM fluctuation.

Let us also remind that the linear theory does not prescribe equal
{\it linear} amplitudes for DM and baryons, but that their ratio is
given by eq.~(\ref{biasl}). Therefore, it becomes also important to
outline that we chose to refer to DM fluctuation amplitudes in the
linear regime. Accordingly, let $\delta_{c,rc}^{(c)}$ and
$\delta_{c,rc}^{(b)}$ be both DM fluctuation amplitudes, defined so
that, if the linear theory yields the former (latter) value at a time
$\tau$, the corresponding DM (baryon) fluctuation has fully
recollapsed at $\tau$.

In Figure \ref{deltac} we plot the $\beta$ dependence of
$\delta_{c,rc}^{(c)}$ and $\delta_{c,rc}^{(b)}\, $, at $z=0$. They
start from equal values for $\beta=0$ and gradually split as $\beta$
increases. In Figure \ref{dcz} we plot the $z$ dependence of
$\delta_{c,rc}^{(c)}$ and $\delta_{c,rc}^{(b)}\, $ for $\beta=0.20$
and 0.05.  These values were computed by starting at $z = 1000$,
taking into account also the radiative component, and using the full
set of equations.

Following the PS approach, the differential mass function then
reads
$$
\psi (M) = 2 {\rho \over M} \int_{\delta_{c,rc}}^\infty d\delta_M
~{d\sigma_M \over dM} \, {d \over d\sigma_M} \left\{ {1 \over
\sqrt{2\pi} \sigma_M} e^{-\delta_M^2 /2 \sigma_M^2} \right\}
$$
\begin{equation}
= \sqrt{2 \over \pi} {\rho \over M}
\int_{\delta_{c,rc}/\sigma_M}^\infty d\nu_M
~{d\nu_M \over dM} \, \nu_M
e^{-\nu_M^2 /2 } ~.~~
\label{ps}
\end{equation}
Here $\nu_M = \delta_M/\sigma_M$ and $\delta_{c,rc}$ can be either
$\delta_{c,rc}^{(c)}$ or $\delta_{c,rc}^{(b)}$ (or any intermediate
value) according to Fig.~\ref{deltac} and \ref{dcz}; a choice shall be
based on the observable to be fitted. 

The ST expression is obtainable from eq.~(\ref{ps}) through the
replacement
\begin{equation}
\nu_M \exp (-\nu_M^2/2) ~~\to~~
{\cal N}'\, {\nu'}_M (1+{\nu'}_M^{-3/5})\, \exp (-{\nu'}_M^2/2)~,
\label{st}
\end{equation}
$$
{\rm with} ~~~{\cal N}'=0.322~,~~~~ {\nu'}^2_M = 0.707\, \nu^2_M~,
$$ 
meant to take into account the effects of non--sphericity in the halo
growth.

In the absence of coupling and baryon--DM segregation, the mass $M$ in
the PS and ST expressions (\ref{ps})--(\ref{st}) {\it is} the mass
originally in the top--hat, which will then be comprised within a
virial radius $R_v$, such that $M/(4\pi/3)\rho R_v^3 = \Delta_v$. 

In the presence of coupling and segregation the situation is more
complex. However, indipendently of the value taken for
$\delta_{c,cr}$, {\it the resulting virialized system will be baryon
depleted}. Different possible $\delta_{c,cr}$'s will correspond to
different depletions, but the final system shall however contain a
smaller fraction of baryons, in respect the background
$\Omega_b/\Omega_c$ ratio.

It is important to distinguish between two effects: (i) DM mass
variation. (ii) The dynamics of gravitational growth. Let $M_i$ and
$M_{vir}$ be the masses, at the initial time and at virialization,
{\it rescaled to the values they will have at $\tau_o$}, so that the
(i) effect is isolated. Then, while
\begin{equation}
M_i = M_{i}^c + M_{i}^b = (\Omega_c/\Omega_m) M_{i} + (\Omega_b/
\Omega_m) M_{i}~,
\end{equation}
so that $ M^c_i / M^b_i= \Omega_c/\Omega_b\, $, in the
decomposition
\begin{equation}
M_{vir} = M_{vir}^c + M_{vir}^b
\end{equation}
it will however be $ M^c_{vir} / M^b_{vir} >  
\Omega_c/\Omega_b\, $.

Let us now duly take into account also the (i) effect and consider the
case when the mass function is set by $\delta_{c,rc}^{(c)}$.  Then,
while $M^c_{vir}/ M^c_{i} = \exp[-C(\phi_{vir}-\phi_i)]$, it will
obviously be $M^b_{vir} < M^b_{i} $: several baryon layers, initially
belonging to the fluctuation, have not yet recollapsed or virialized.

Let us then consider the mass function set by $\delta_{c,rc}^{(b)}$.
In this case it is $M^b_{vir} = M^b_{i}$, but it will be $M^c_{vir}/
M^c_{i} > \exp[-C(\phi_{vir}-\phi_i)]$. The extra DM mass is due to
those layers, initially external to the fluctuation, first compressed
and then conveyed inside the virialization radius, together with the
baryons previously outflown from the DM bulk.

For any $\delta_{c,rc}$ in the $\delta_{c,rc}^{(c)}$--$\delta
_{c,rc}^{(b)}$ interval, some baryon layers will still be out and some
extra DM will have been conveyed inside the virial radius by the fall
out of outflown baryons. Hence, it will however be 
$ M^c_{vir}\exp[-C(\phi_o - \phi_{vir}] ~/M^b_{vir} > \Omega_c/\Omega_b\, $.

In Figure \ref{mdelta} we plot this ratio, as a function of
$\delta_{c,rc}$, in the $\delta_{c,rc}^{(c)}$--$\delta_{c,rc}^{(b)}$
interval. The plot shows that, after a fast decrease, the ratio tends
to a steady value, however exceeding the background ratio. The curves
shown in this plot depend on the assumed (top--hat) shape for the
primeval fluctuation, but similar curves would hold for any initial
shape.

A prediction of cDE theories, therefore, is that $\Omega_c/\Omega_b$,
measured in any virialized structure, exceeds the background ratio.
The excess is greater for structures where only DM has virialized.
They might be characterized by an apparent disorder in the baryon
component, still unsettled in virial equilibrium while, {\it e.g.}, a
lensing analysis would show that they are safely bound systems.

\begin{figure}
\begin{center}
\vskip -0.5truecm
\includegraphics*[width=10cm]{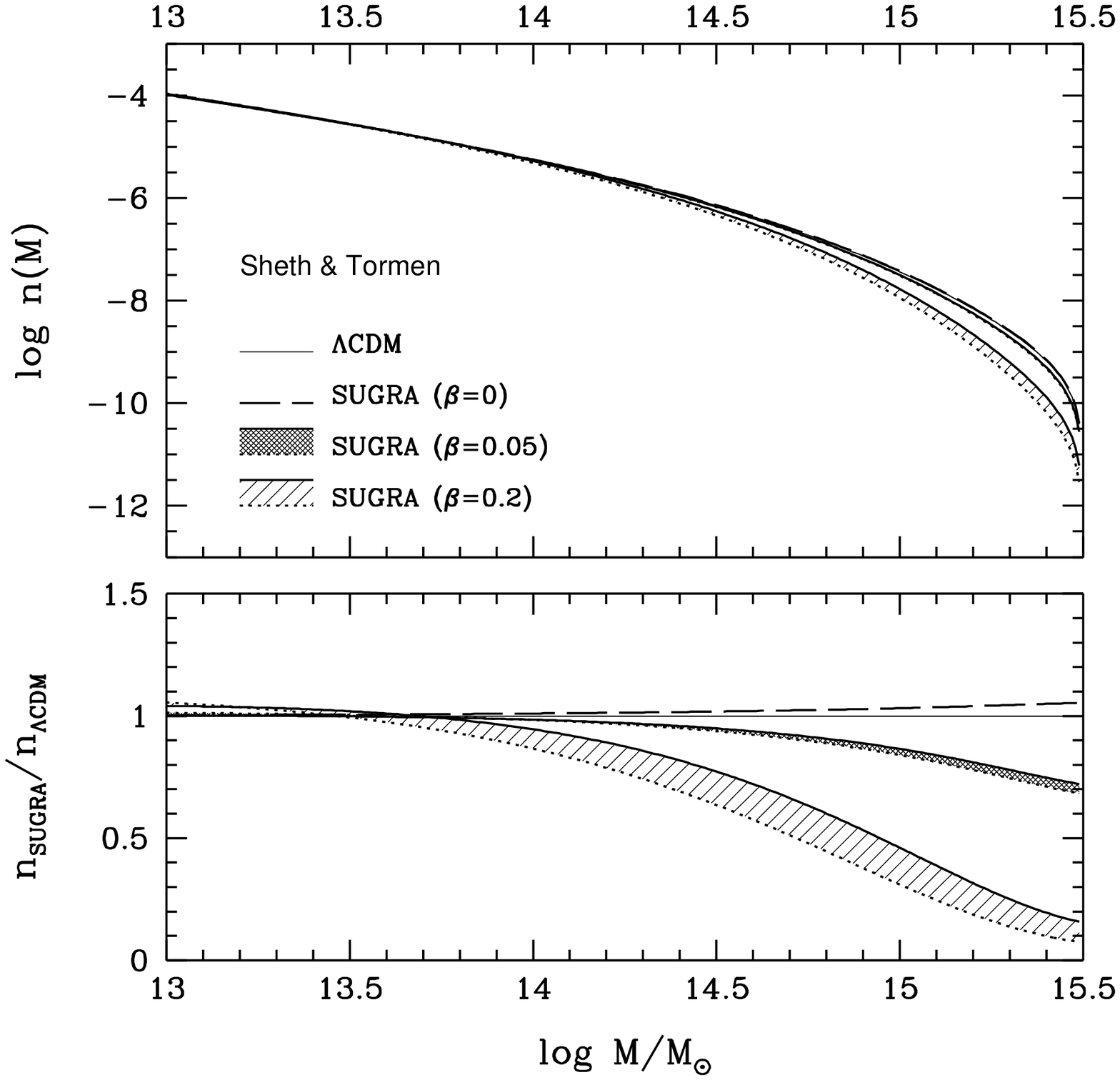}
\end{center}
\vskip -1.3truecm
\caption{Cluster per (Mpc $h^{-1})^3$, above mass $M$ at $z=0$,
 obtained from ST expression. Four models are
considered, with equal $\Omega$'s and $h$: $\Lambda$CDM, uncoupled
SUGRA with $\Lambda = 100\, $MeV, and two cDE with different
$\beta$'s.  The dashed areas are limited by the mass functions worked
out for $\delta_{c,rc}^{(c)}$ or $\delta_{c,rc}^{(b)}$ in cDE models
(see text).  
}
\label{mfz0ST}
\begin{center}
\includegraphics*[width=10cm]{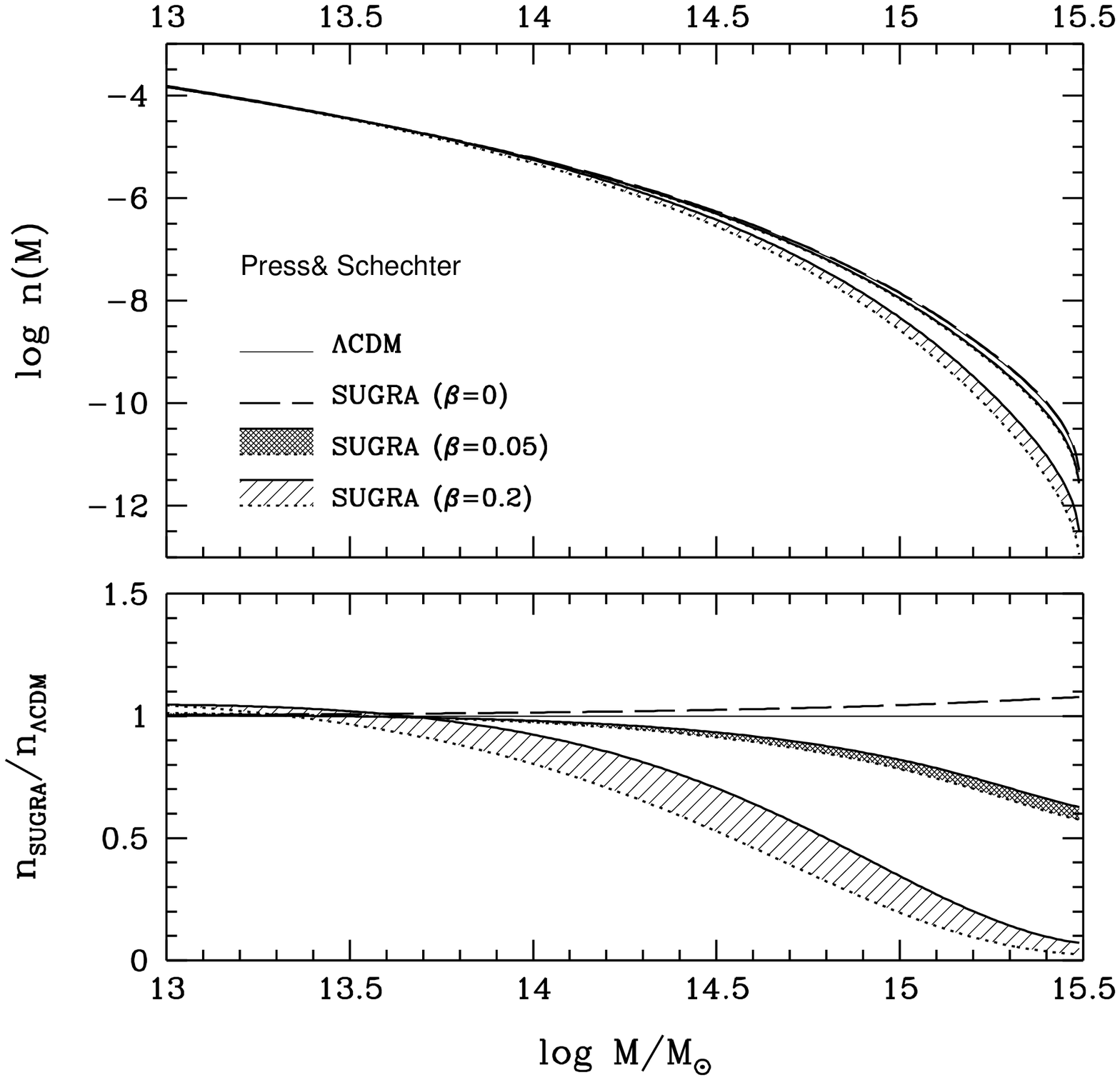}
\end{center}
\vskip -1.3truecm
\caption{The same as Fig.~\ref{mfz0ST}, using the original PS expression. 
This figure shows just slight quantitative shifts.
}
\label{mfz0}
\end{figure}
%
\begin{figure}
\begin{center}
\includegraphics*[width=10cm]{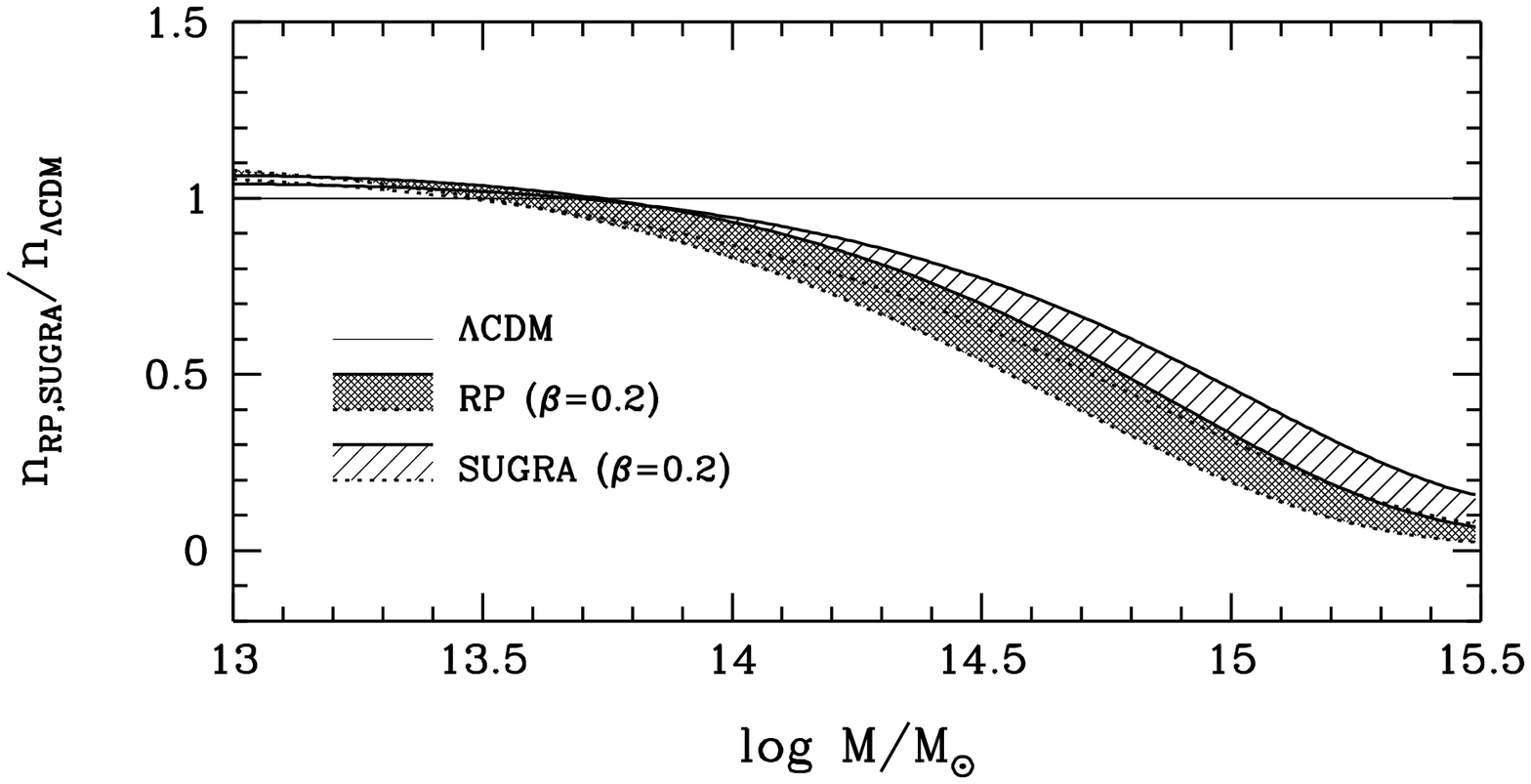}
\end{center}
\vskip -4.5truecm
\caption{A comparison of SUGRA and RP results at $z=0$ using the ST expression.
The lower panel of Fig.~\ref{mfz0ST} is reproduced, neglecting dDE and
$\beta=0.05$ cases and adding RP results (heavy dashed).
}
\label{mfzRP0}
\end{figure}

We shall now plot mass functions obtained using either
$\delta_{c,rc}^{(c)}$ or $\delta_{c,rc}^{(b)}$. We expect actual
measures to yield a value comprised in this interval and, however,
closer to the $\delta_{c,rc}^{(c)}$ curve when baryon stripping is
stronger. In Figure \ref{mfz0ST} we plot the integral mass functions
$n(>$$M) = \int_M^\infty dM'\, \psi(M')$ obtained through ST 
expressions (\ref{ps})--(\ref{st}). 
Let us remind that large differences
between models were never found in mass functions at $z=0$, because of
DE nature. The upper panel of each figure shows the mass function in the usual
fashion, as often plotted to fit data or simulations. In the lower
panel we plot the ratio between expected halo numbers for each model
and $\Lambda$CDM. This confirms the small shifts between $\Lambda$CDM
and dDE cosmologies, yielding just a slight excess, $\sim 10\, \%$, on
the very large cluster scale, where observed clusters are a few units.

Discrepancies can be more relevant between $\Lambda$CDM and cDE, whose
effective mass function should however lie inside the dashed areas,
limited by the function obtained by integrating from
$\delta_{c,rc}^{(c)}$ or $\delta_{c,rc}^{(b)}\, $. The plots show a
shortage of larger clusters. For $\beta = 0.20$, they are half of
$\Lambda$CDM at $\sim 3 \cdot 10^{14} h^{-1} M_\odot$ and really just
a few above some $10^{15} h^{-1} M_\odot\, $. Any realistic mass
function, laying in the dashed area, can be falsified by samples just
slightly richer than those now available.

For $\beta = 0.05$ the shift is smaller, hardly reaching $20 \, \%$,
still in the direction opposite to dDE. Here we meet what appears to
be a widespread feature of cDE models: the discrepancy of dDE from
$\Lambda$CDM is partially or totally erased even by a fairly small
DM--DE coupling, and many cDE predictions lay on the opposite side of
$\Lambda$CDM, in respect to dDE. If a $\Lambda$CDM model is then used
to fit galaxy or cluster data arising in a cDE cosmology, we expect 
cluster data to yield a best--fit $\sigma_8$ value smaller than the
one obtained by fitting galaxy data.

The integral mass functions obtained through the original PS
expression are shown in Fig.~\ref{mfz0}. This figure shows just slight
quantitative shifts.  In the cases illustrated by the next figures,
results from PS expression will be therefore omitted.

Figure \ref{mfzRP0} compares RP results with SUGRA, at $z=0$.
Quantitative differences exist, when the self--interaction
potential is changed, but most physical aspects are the same.

\begin{figure}
\begin{center}
\includegraphics*[width=10cm]{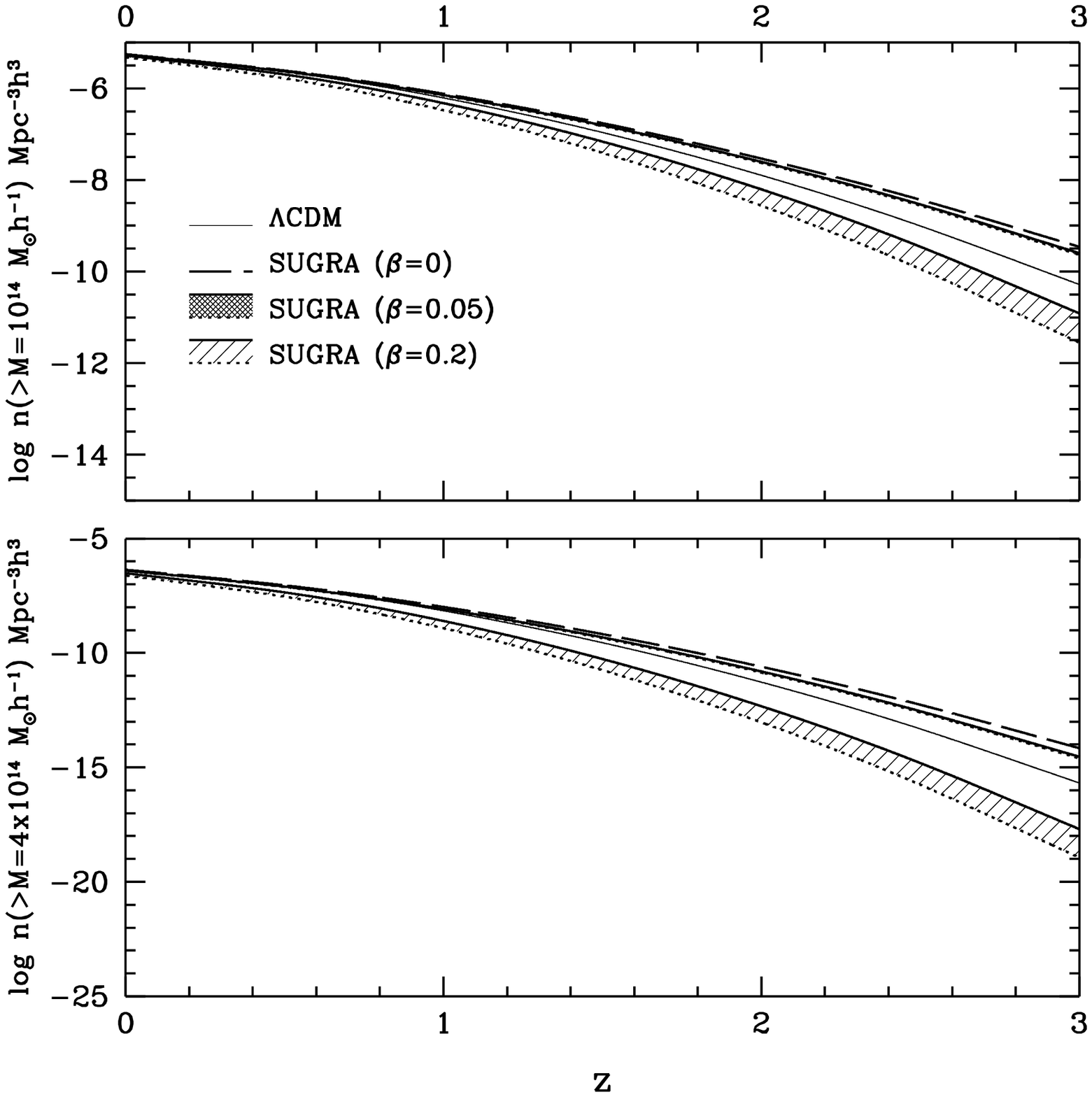}
\end{center}
\vskip -1.3truecm
\caption{Cluster per (Mpc $h^{-1})^3$, above mass $M$ at different $z$'s, in a fixed
comoving volume, obtained from ST expression. 
Thick solid and dotted lines refer to cDE models.
The dashed (thinner solid) line refers to dDE ($\Lambda$CDM).  }
\label{mfz}
\end{figure}

\begin{figure}
\begin{center}
\includegraphics*[width=11cm]{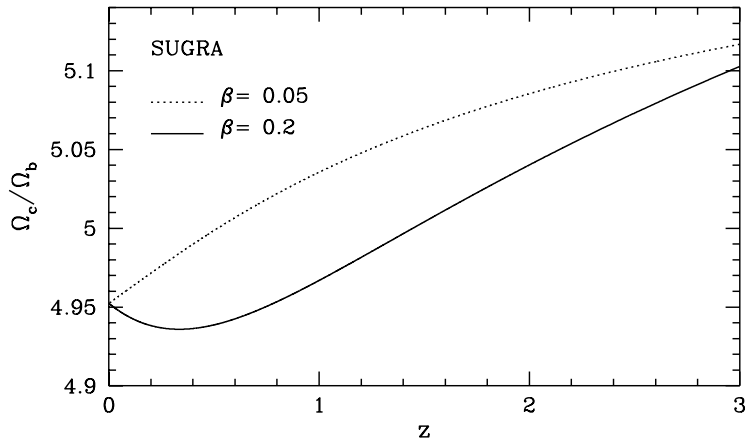}
\end{center}
\vskip -5.truecm
\caption{Evolution of the DM/baryon background ratio, due to the
dynamics of the $\phi$ field.
}
\label{omegarap}
\end{figure}

\begin{figure}
\begin{center}
\includegraphics*[width=10cm]{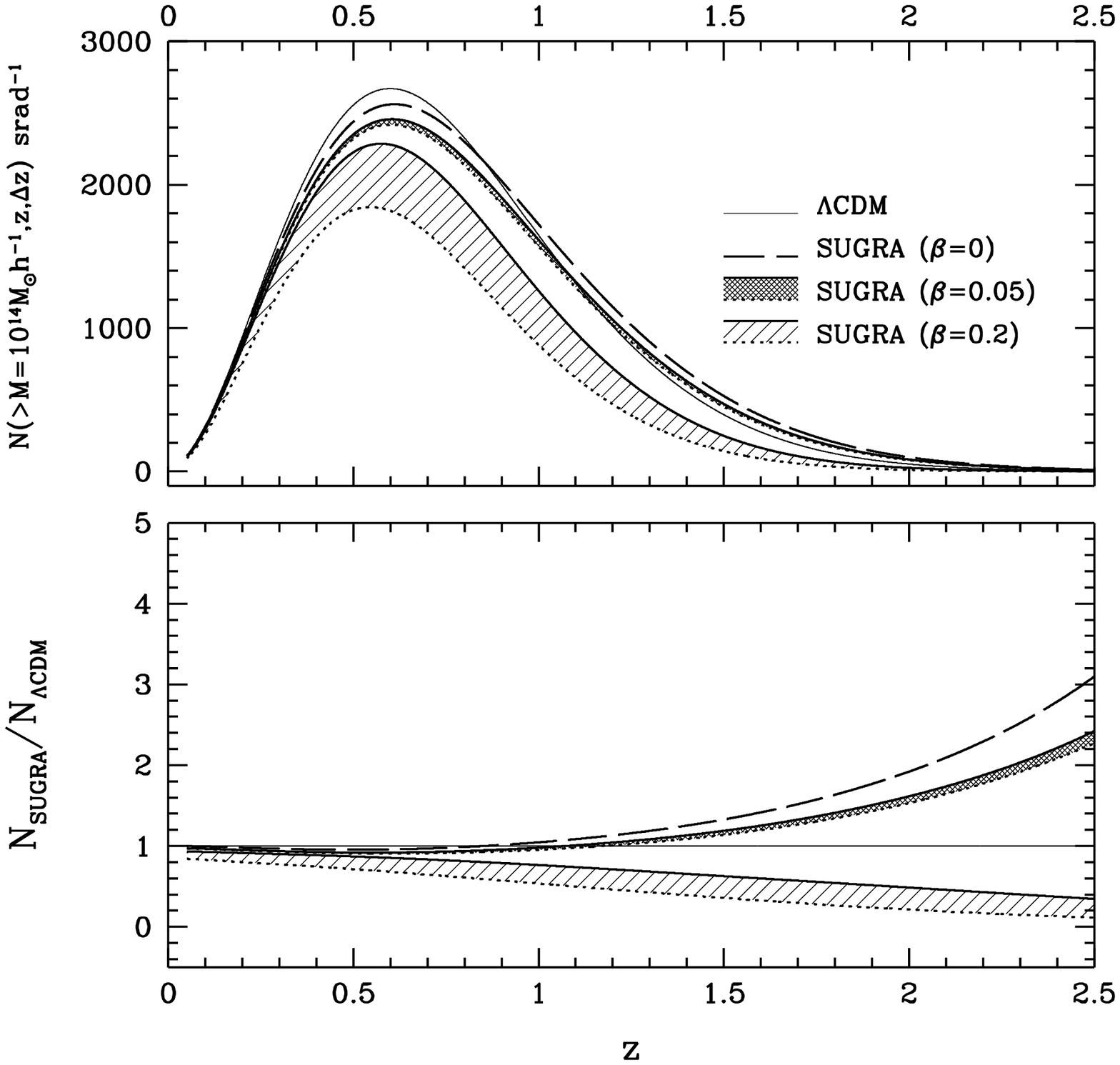}
\end{center}
\vskip -1.5truecm
\caption{Number of clusters with $M>10^{14}h^{-1} M_\odot\, $ in a
fixed solid angle and redshift interval.
Lines as in Fig.~\ref{mfz}.
}
\label{cmfST}
\begin{center}
\includegraphics*[width=10cm]{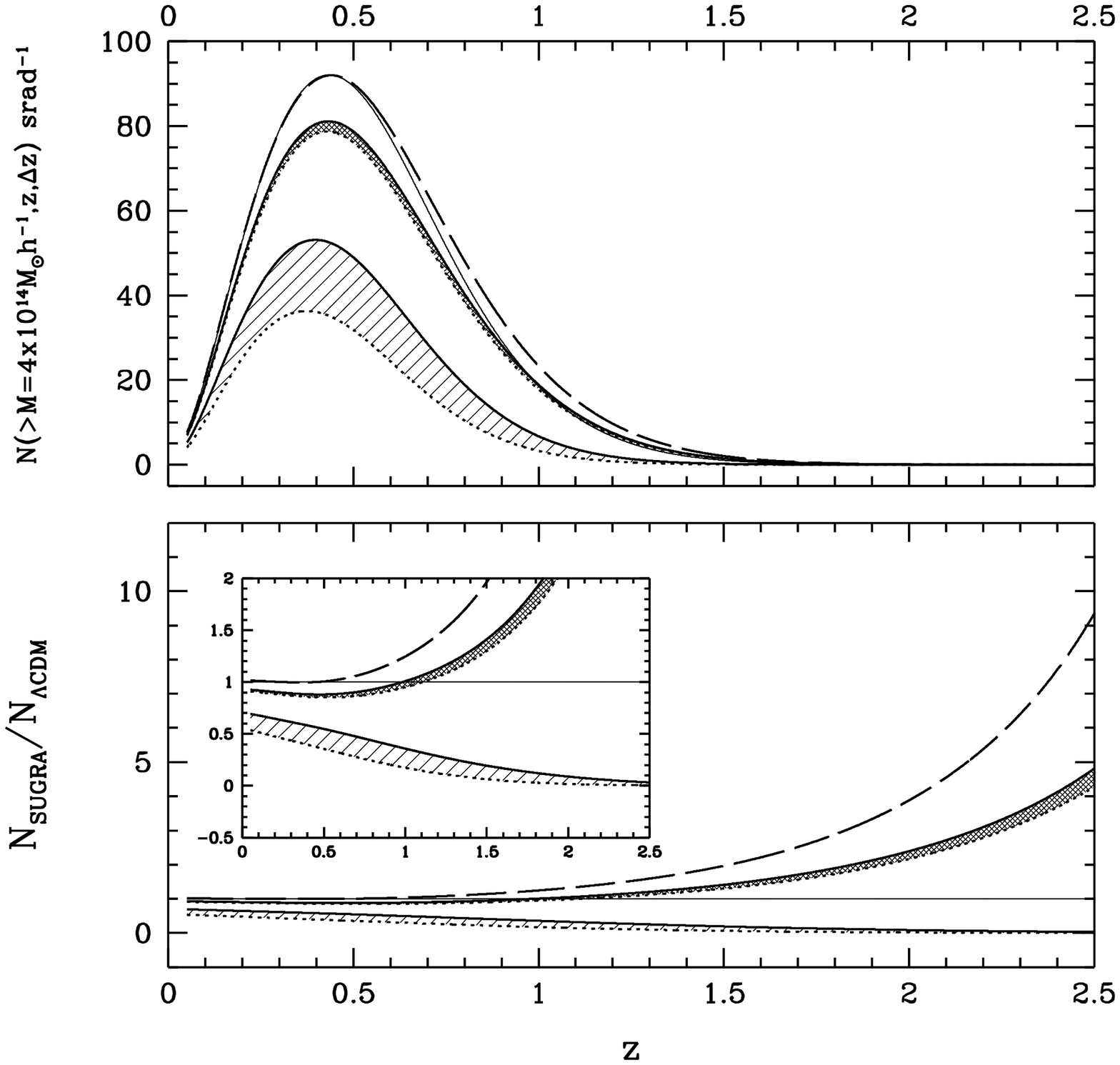}
\end{center}
\vskip -1.5truecm
\caption{Number of clusters above $4 \cdot 10^{14}h^{-1} M_\odot\, $.
Lines and comments as for Fig.~\ref{cmfST}.
}
\label{cmf4ST}
\end{figure}

\begin{figure}
\begin{center}
\vskip -0.5truecm
\includegraphics*[width=10cm]{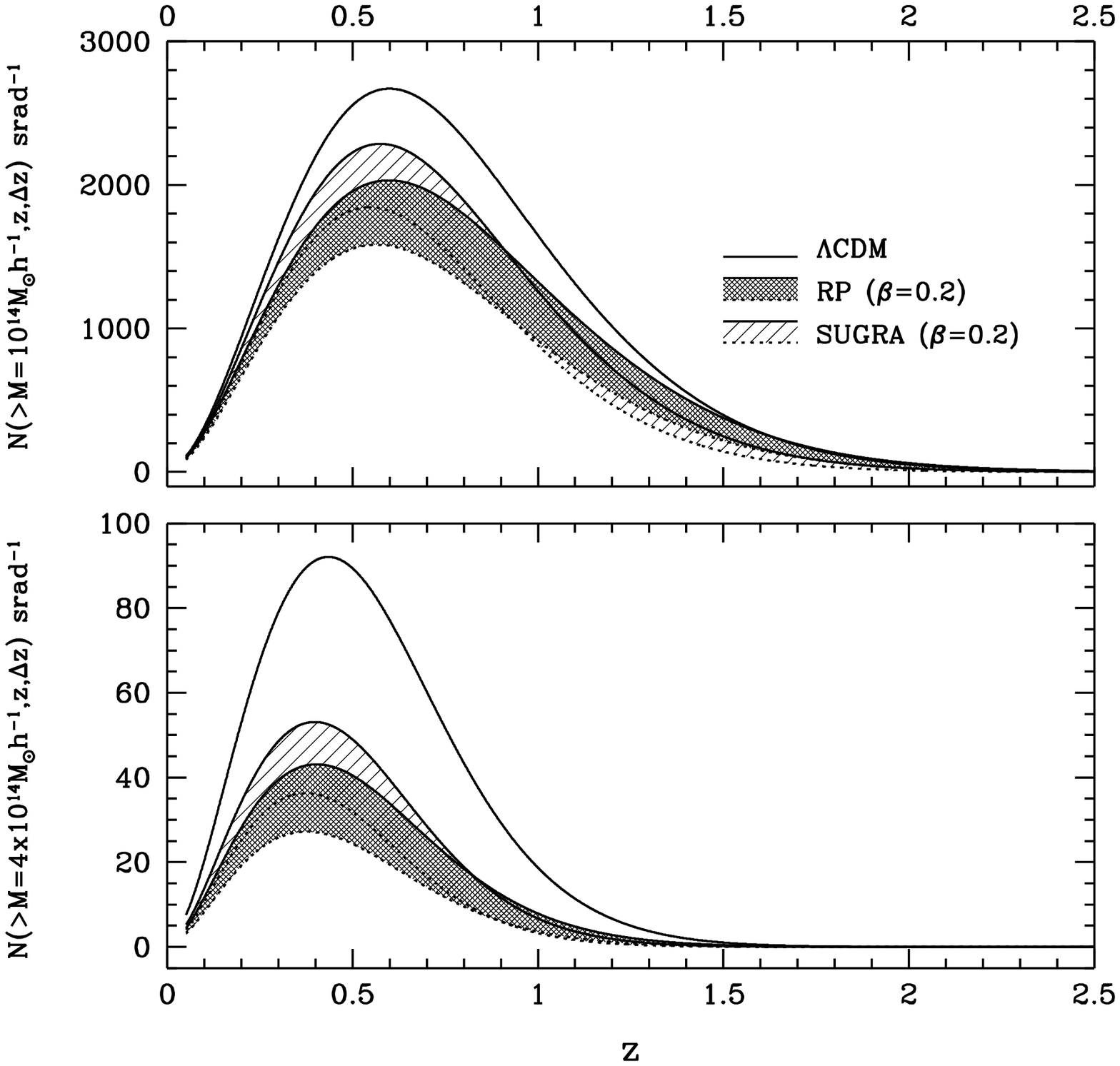}
\end{center}
\vskip -1.3truecm
\caption{A comparison between SUGRA and RP results.
Here the top panels of Figs.~\ref{cmfST} and \ref{cmf4ST} are reproduced,
omitting the $\beta = 0.05$ case and showing both SUGRA (dashed)
and RP (heavy dashed) number intervals. The number distribution 
of $\Lambda$CDM is also given (solid line).
}
\label{cmfRP4}
\end{figure}

In Figure \ref{mfz} the redshift dependence of the expected cluster
numbers {\it in a comoving volume} is plotted against the redshift
$z$, for $M = 10^{14}$ and $4 \cdot 10^{14} h^{-1} M_\odot$. 
As usual, the mass considered is the total cluster mass. As already widely
outlined, the DM/baryon ratio in these masses however exceeds
the background $\Omega_b/\Omega_c$ ratio; the spread of the function
corresponds to the spread of possible baryon/DM ratios. In top of
that, however, one must also remind that the very background
$\Omega_c/\Omega_b$ ratio varies with redshift, because of the
evolution of the $\phi$ field. Hence, clusters observed at high $z$,
in average, shall be however baryon poorer than present time clusters.
The $z$ dependence of the background $\Omega_c/\Omega_b$ ratio, for
the model considered in this work, is plotted in Figure \ref{omegarap}.

Plots similar to Figure \ref{mfz} are often used to assert the
possibility to discriminate between models and this plot is however
significant to compare cDE with former results for other
cosmologies. According to \cite{sole}, however, the discriminatory
capacity of this observable can only be tested by plotting cluster
numbers per solid angle and redshift interval, which also includes
geometrical effects, often partially erasing dynamical effects.

In the upper panels of Figures \ref{cmfST} and \ref{cmf4ST}
cluster numbers per solid angle and redshift interval are plotted. In the
lower panels we plot the ratios between each SUGRA model and
$\Lambda$CDM mass functions. We consider again the mass scales
$10^{14}$ and $4 \cdot 10^{14} h^{-1} M_\odot$; for the latter mass we
provide a magnified box to follow the expected low--$z$
behavior. Notice, in particular, that the high--$z$ behaviors for
$\beta = 0.05 $ or 0.2, for these mass scales, lay on the opposite
sides of $\Lambda$CDM. The box in Fig. \ref{cmf4ST} 
show how we pass from numbers smaller than $\Lambda$CDM to greater numbers, for $\beta
= 0.05$, at a redshift $z \simeq 0.7\, $.

As is obvious, a better discrimination is attained for high masses or
deep redshifts. However, according to Fig.~\ref{cmf4ST}, a sample
including a few dozens clusters of mass $>4 \cdot 10^{14} h^{-1}
M_\odot$ at $z > 0.5$ would already bear a significant discriminatory
power.

Altogether we see that, (i) when passing from $\Lambda$CDM to
uncoupled SUGRA, the high--$z$ cluster number is expected to be
greater. (ii) When coupling is added, the cluster number excess is
reduced and the $\Lambda$CDM behavior is reapproached. (iii) A
coupling $\beta = 0.05$ may still yield result on the upper side of
$\Lambda$CDM, while $\beta = 0.2$ displaces the expected behavior well
below $\Lambda$CDM. The $\Lambda$CDM behavior is approsimately met for
$\beta = 0.1$.

These behaviors arise, first of all, from the different evolution of
$\rho_{DE}$ at high $z$. In $\Lambda$CDM models, $\rho_{DE}$ simply
keeps constant, while $\rho_m \propto (1+z)^3$, and therefore its
relevance rapidly fades. In dDE models, instead, $\rho_{DE}$ increases
with $z$, although at a smaller rate than $\rho_m$. To obtain the same
amount of clusters at $z=0$ requires that they have existed since
earlier.

When coupling is added, however, gravitation is boosted by the $\Phi$
field. In the newtonian language, this translates into a greater
gravity constant ($G^*$) and growing masses for DM particles. This
speeds up cluster formation and, when $\beta $ increases, less
clusters are needed at high $z$, to meet their present numbers.

Finally, in Figure \ref{cmfRP4} we provide a comparison between
the PS--ST mass function obtainable for RP and SUGRA potentials. Only the
case $\beta=0.2$ is considered. A RP potential slightly strengthens
the effects already seen for SUGRA; as a matter of fact, however, 
in most cases we find just small quantitative shifts.

\section{Conclusions}
In this work we aimed at predicting the cluster mass function in cDE
cosmologies, by using the solution of the equations ruling the
spherical growth of top--hat fluctuations in ST (or PS) expressions.
The effectiveness of ST expressions has been widely verified for SCDM,
$\Lambda$CDM and 0CDM models. Also in simulations of models with
dynamical (uncoupled) DE \cite{klypin2003}, ST expressions provide a
fair fit of numerical outputs.

The main finding of this study is the significant baryon--DM
segregation, which has multiple effects. If cluster numbers are
measured from their gravitational effects, {\it e.g.}, by using
lensing data, a PS--ST approach yields simple predictions. When
cluster numbers are measured through other observables, we predict a
number range, inside which observations should lie. The actual amount
of objects, inside these ranges, is determined by a number of effects,
that a PS--ST approximation cannot describe.

An important example of such effects is the possible stripping of
outer layers, in close encounters, which will mostly act on the baryon
component. For a rather small coupling as $\beta = 0.2$, up to $40 \,
\%$ of the baryons belonging to the initial fluctuation could be
stripped in this way and, even for the tiny coupling set by $\beta =
0.05$, 10$\, \%$ of baryons could be easily stripped.

If cluster data are obtained from galaxy counts or hot gas features,
they will exhibit the residual baryon amount. Indipendently of the
baryon loss, which depends on the individual cluster history, cDE
theories predict that the background $\Omega_c/\Omega_b$ ratio however
increases with redshift. However, in top of that, the DM/baryon ratio
measured in any virialized structure, exceeds the background ratio at
the redshift where it is observed and is expected to exhibit
significant variations in different systems, being smaller in larger
systems, in average.

It must be however outlined that the final baryon/DM ratio, in any
galaxy cluster at any redshift, {\it even in the absence of any
stripping effect}, is expected to by smaller than the background
$\Omega_b/\Omega_c$ ratio at that redshift. In this paper dedicated to
a technical analysis of cDE mass functions we refrain from discussing
this feature in further detail, although relating it with the apparent
baryon shortage in clusters, (see, {\it e.g.}, \cite{allen}) seems
suggestive.

Furthermore, when cluster data are obtained through the hot gas
behavior, a complex interplay between baryons and potential well is
expected. Once again, however, the model used for PS--ST estimates
seems unsuitable to provide quantitative predictions, but {\it
anomalies} in the temperature--luminosity relations are expected.

A PS--ST analysis allows however to formulate further predictions,
besides those concerning the $\Omega_b/\Omega_c$ ratio in galaxy
clusters and in galaxies. They concern the cluster mass function and
its evolution.

No large differences between models were ever found in the mass
functions at $z=0$, because of DE nature: just a slight excess, $\sim
10\, \%$, on the very large cluster scale, where observed clusters are
a few units, was found in dDE, in respect to a $\Lambda$CDM cosmology.

Discrepancies can be more relevant between $\Lambda$CDM and cDE, where
a {\it shortage} of larger clusters is predicted. For $\beta = 0.20$,
they are half of $\Lambda$CDM at $\sim 3 \cdot 10^{14} h^{-1} M_\odot$
and less than 20$\, \%$ above a few $10^{15} h^{-1} M_\odot\, $.  Such
strong shortage can be falsified by samples just slightly richer than
those now available. For $\beta = 0.05$ the shift is smaller, hardly
reaching $20 \, \%$, but still in the direction opposite to dDE.

This is a widespread feature of cDE models: the discrepancy of dDE
from $\Lambda$CDM is partially or totally erased even by a fairly
small DM--DE coupling, and many cDE predictions lay on the opposite
side of $\Lambda$CDM, in respect to dDE. Therefore, if a $\Lambda$CDM
model (or any uncoupled model) is used to fit galaxy or cluster data
arising in a cDE cosmology, we expect that cluster data may yield a
smaller $\sigma_8$, in comparison to the one worked out from other
data sets.

Turning to the evolutionary predictions, we expect an evolution {\it
faster} than $\Lambda$CDM for any coupling $\beta > 0.1$, again the
opposite of what we expect in uncoupled dDE.

Most quantitative results given in this paper are worked out by
assuming that the scalar field self--interaction potential is SUGRA,
with $\Lambda = 100\, $GeV. For the sake of comparison, in a few
plots, results obtained for a RP potential are also shown. It should
be reminded that, while the former potential predicts a linear
behavior consistent with observations, the latter one can be fitted
with linear observables only for quite low values of $\Lambda$, much
below the one considered here; we selected it just to provide a direct
comparison tool, so allowing us to conclude that a different
self--interacion potential may cause quantitative shifts up to some
10$\, \%$, but hardly affect our general conclusions.

Altogether, there can be no doubt that cDE cosmologies open new
prespectives for the solution of those problems where baryon--DM
segregation due to hydrodynamics is apparently insufficient to explain
observed features. These problems may range from the shortage of
galaxy satellites in the local group, to galactic disk formation, up
to the $L$ vs $T$ relation in large clusters. Stating how coupling can
affect these and similar questions, by using a PS--ST approach, is
hard.  This calls for detailed n--body and hydrodynamical simulations
of cDE cosmologies.

\begin{acknowledgments}

\section{Acknowledgments}
Luca Amendola, Andrea Macci\`o and Loris Colombo are gratefully
thanked for their comments on this work.

\end{acknowledgments}

{}

\vfill\eject
\noindent
{\bf Appendix 1. The newtonian regime}

\vglue .4truecm
\noindent
In the presence of inhomogeneities, the metric can read
\begin{equation}
ds^{2}=a^2(\tau)[-(1+2\psi )d\tau ^{2}+(1-2\psi )dx_{i}dx^{i}],
\end{equation}
provided that no anisotropic stresses are considered, $\psi $ being
the gravitational potential in the Newtonian gauge. Let us describe
DE field fluctuations $\delta \phi $ through
\begin{equation}
\varphi = (4\pi/3)^{1/2}\, (\delta \phi/m_p) 
\label{varphi}
\end{equation}
and expand fluctuations in components of wavenumber {\bf k}; let also
be $\lambda ={\mathcal{H}}/k$. Let then be
\begin{equation}
f = \phi^{-1} \sqrt{3/16\pi G} ~\ln(V/V_o)
,~~
f_{1}=\phi {\frac{df}{d\phi }}+f,~~ 
f_{2}= \phi {\frac{df}{d\phi }}+2f+f_{1}~;
\end{equation}
$V_o$ being a reference value of the potential. It is also useful to
define $Y^{2}=8\pi GV(\phi)\, a^2 /3 {\mathcal{H}}^{2}$.  The
equations ruling the evolution of the $\varphi$ field and gravity,
keeping just the lowest order terms in $\lambda$, as is needed to
obtain their Newtonian limit, then read
\begin{equation}
\psi = -{3 \over 2} \lambda^2 (\Omega_b \delta_b +
\Omega_c \delta_c + 6X \varphi + 2X \varphi'- 2 Y^2 f_1\, \varphi )~,~~
\psi' = 3 x \varphi - \psi ~,~~~~~~~~
\label{poten1}
\end{equation}
\begin{equation}
\varphi ''+\big (2+\frac{ { \cal H}^{'}}{ {\cal H}}\big )\varphi '+
\lambda ^{-2}\varphi -12X\varphi +4\psi X+2Y^{2}(f_{2}\varphi -f_{1}\psi )=
\beta \Omega _{c}(\delta _{c}+2\psi )~;
\label{phi2}
\end{equation}
let us remind that $X$ is defined in eq.~(\ref{ics}) and notice that,
if DE kinetic (and/or potential) energy substantially contributes to
the expansion source, $X$ (and/or $Y$) is $\mathcal{O}$(1).

In the Newtonian limit, $\varphi $ derivatives shall be neglected, the
oscillations of $\varphi $ and the potential term $f_{2}Y^{2}\varphi $
should be averaged out, by requiring that $\lambda <<(f_{2}Y)^{-1}$,
and, in eq.~(\ref{phi2}), the metric potential $\psi $ ($\propto
\lambda^2$) can also be neglected. Then,
eqs.~(\ref{poten1})--(\ref{phi2}) become
\begin{equation}
\psi =-\frac{3}{2}\lambda ^{2}(\Omega _{b}\delta _{b}+\Omega _{c}\delta _{c}),\
~,~~
\lambda ^{-2}\varphi \simeq \beta \Omega _{c}\delta _{c}.\label{approx2}~.
\label{approx1}
\end{equation}
Baryon and DM density fluctuations are then ruled by the
eqs.~(\ref{delta2})--(\ref{ar2}), derived from the stress--energy {\it
pseudo}--conservation $T_{\nu ;\mu }^{\mu }=0$. Thereinside, $'$
yields differentiation with respect to $\ln a$. [Let us then notice
that, taking $\Omega_b << \Omega_c$, putting $\delta_c \propto e^{\int
\mu(\alpha) d \alpha}$ and $\delta_b = b \delta_c$ with $b={\it
cost}$, eqs.~(\ref{delta2}), allow us to obtain the bias factor
(\ref{biasl})].

The acceleration of a DM or baryon particle of mass $m_{c,b}$ can then
be derived from
eqs.~(\ref{delta2})--(\ref{ar2}). Let it be
in the void, at a distance $r$ from the origin, where a DM (or baryon)
particle of mass $M_{c}$ (or $M_{b}$) is set, and let us remind that,
while $\bar \rho_b \propto a^{-3}$, it is
\begin{equation}
\bar \rho_c = \bar \rho_{oc} a^{-3}
e^{-C (\phi -\phi_0)},~~ \rho_{M_c} = M_{oc} a^{-3} e^{-C (\phi -\phi_0)}
\delta(0),
\label{densi}
\end{equation}
because of the DE--DM coupling (the subscript $o$ indicates values at
the present time $\tau_{o}$; let also be $a_{o}=1$). We can then
assign to each DM particle a varying mass $M_c (\phi)=M_{oc}e^{-C
(\phi -\phi_0)}$

Then, owing to eq.~(\ref{densi}), and assuming that the density of the
particle widely exceeds the background density, it is
\begin{equation}
\Omega_c \delta_c = {\rho_{M_{c}} - \bar \rho _{c} \over \rho _{cr}}
={ 8 \pi G \over 3 {\mathcal{H}}^2 a} M_{c}(\phi) \delta (0) ~,~~
\Omega_b \delta_b = 
{\rho_{M_{b}}- \bar \rho_b \over \rho_{cr}}
={8\pi G \over 3{\mathcal{H}}^2 a}  M_b \delta (0),
\end{equation}
($\rho_{cr}$ is the critical density and $\delta $ is the Dirac
distribution). Reminding that $\nabla \cdot {\bf v}_{c,b}=
\theta_{c,b} {\mathcal{H}}$ and using the ordinary (not conformal)
time, eq.~(\ref{delta2}) yields
\begin{equation}
\nabla \cdot \dot {\bf v}_c=
- H (1-2\beta X)\, \nabla \cdot {\bf v}_c - 4\pi G a^{-2}
\left( \gamma M_{c}(\phi)  + M_{b} \right) \delta (0)
\label{velocity2}
\end{equation}
(dots yield differentiation in respect to ordinary time and
$H=\dot{a}/a$). Taking into account that the acceleration is radial,
as the attracting particles are in the origin, it will be
\[
\int d^{3}r \, \nabla \cdot {\dot{\textbf {v}}}
=4\pi \int dr~ {d(r^2 \dot{v})/dr}=4\pi r^{2} \dot{v}.
\]
Accordingly, the radial acceleration of a DM particle is
\begin{equation}
\dot{v}_{c}=-(1-2\beta X)H{\textbf {v}}_{c}\cdot 
{\textbf {n}}-\frac{G^{*}M_{c}(\phi)}{R^{2}}-\frac{GM_{b}}{R^{2}},
\label{finalc}
\end{equation}
($\textbf {n}$ is a unit vector in the radial direction; $R=ar$). 

By following a similar procedure for baryons, we obtain:
\begin{equation}
\dot{v}_{b}=-H{\textbf {v}}_{b}\cdot 
{\textbf {n}}-{\frac{GM_{c}(\phi)}
{R^{2}}}-{\frac{GM_{b}}{R^{2}}}
\label{finalb}
\end{equation}
In the presence of full spherical symmetry, $\bf {v}_{b,c} \cdot 
{n}$ $= v_{b,c}$; being then $b=r_b$ and $c=r_c$, eqs.~(\ref{shells})
immediately follow.

For the sake of completeness, here below, we write also the equations
for the physical DM and baryons radii which easily follow from
eqs.~(\ref{finalc}) and (\ref{finalb}).  While for baryons we have the
usual Friedman-like equations
\begin{equation}
\ddot R_b^n = - {4\pi \over 3}G ~[\rho_c + \rho_b + \rho_{DE} (1+3w)]~R_b^n ~,
\end{equation}
for DM we have
\begin{equation}
  \ddot R_c^n = C \dot \phi \dot R_c^n - C \dot \phi H R_c^n
- {4\pi \over 3}G ~[\bar \rho_c + \gamma (\rho_c - \bar \rho_c)~ 
+ \rho_b + \rho_{DE} (1+3w)]~R_c^n.
\end{equation}

\end{document}